\documentclass[aps,pra,twocolumn,superscriptaddress]{revtex4-1}

\usepackage[T1]{fontenc}
\usepackage[latin9]{inputenc}
\usepackage{amsmath}
\usepackage{amssymb}
\usepackage{graphicx}
\usepackage{hyperref}
\usepackage{natbib}
\usepackage{esint}
\usepackage{times}
\usepackage{helvet}
\usepackage{braket}
\usepackage{bbm}
\usepackage{xcolor}
\usepackage{enumitem}
\usepackage[toc,page,titletoc]{appendix}

\makeatletter
\newcommand{\JILA}{JILA, NIST and Department of Physics, University of Colorado, Boulder, CO 80309, USA}
\newcommand{\CTQM}{Center for Theory of Quantum Matter, University of Colorado, Boulder, CO 80309, USA}

\newcommand{\Toronto}{Department of Physics and CQIQC, University of Toronto, Ontario M5S~1A7, Canada}

\begin{document}

\title{Quantum computation toolbox for decoherence-free qubits using multi-band alkali atoms}
\date{\today}

\author{Mikhail Mamaev}
\email{mikhail.mamaev@colorado.edu}
\affiliation{\JILA}
\affiliation{\CTQM}
\author{Joseph H. Thywissen}
\affiliation{\Toronto}

\author{Ana Maria Rey}
\affiliation{\JILA}
\affiliation{\CTQM}

\begin{abstract}
{We introduce protocols for designing and manipulating qubits with ultracold alkali atoms in 3D optical lattices. These qubits are formed from two-atom spin superposition states that create a decoherence-free subspace immune to stray magnetic fields, dramatically improving coherence times while still enjoying the single-site addressability and Feshbach resonance control of state-of-the-art alkali atom systems. Our protocol requires no continuous driving or spin-dependent potentials, and instead relies upon the population of a higher motional band to realize naturally tunable in-site exchange and cross-site superexchange interactions. As a proof-of-principle example of their utility for entanglement generation for quantum computation, we show the cross-site superexchange  interactions can be used to engineer 1D cluster states. Explicit protocols for experimental preparation and manipulation of the qubits are also discussed, as well as methods for measuring more complex quantities such as out-of-time-ordered correlation functions (OTOCs).}
\end{abstract}
\maketitle

\section{Introduction}

The field of ultracold atomic physics in optical lattices has seen tremendous recent growth in its experimental implementations. There are many systems which feature unprecedented levels of cleanliness, environmental isolation, coherence time, and single-site addressability. These capabilities, especially the latter, have motivated recent experiments to apply optical lattice systems to the more ambitious goals of site-resolved quantum simulation~\cite{bloch2017seminalSimulation} and quantum computation~\cite{bloch2008seminalComputation}. An ultracold atomic system realizing a quantum computer can offer the aforementioned benefits of coherence and isolation together with improved scalability, due to the number of atoms these systems can load simultaneously.

The prospect of quantum computing with ultracold neutral atoms has been explored in many contexts. These include the use of collisional gates~\cite{calarco2000collisional,mandel2003collisional}, collective states via cavity QED~\cite{pellizzari1995collective} or dipole blockade of atomic ensembles~\cite{lukin2001collective}, qubits encoded in vibrational atomic states~\cite{eckert2002vibrational,charron2006vibrational}, spin-dependent lattices~\cite{daley2008alkaliEarth}, exchange~\cite{anderlini2007exchange,kaufman2015exchange} or spin-orbit coupled driving~\cite{mamaev2019cluster}, optical tweezers~\cite{weitenberg2011tweezers,pagano2019tweezerComputing}, Rydberg atoms~\cite{molmer2010rydberg,maller2015rydberg,saffman2016rydberg} and several others~\cite{negretti2011review}. However, none of these implementations or techniques can yet claim both single-site resolution and high-quality manipulation (including entanglement generation) in macroscopically-sized arrays. In this work we propose to use fermionic alkali atoms such as $^{40}\mathrm{K}$ in 3D optical lattices to do initialization, state manipulation and entanglement generation with logical qubits encoded in states generated by two atoms in two motional bands. While alkali atoms have been avoided for quantum computation in the past due to their vulnerability to stray magnetic fields, our logical qubits are designed to live in a decoherence-free subspace that resists such unwanted effects, leading to much longer coherence times~\cite{lidar1998decoherenceFree,kielpinski2001decoherenceFree}. With this roadblock removed, we are able to take advantage of the tunability of interactions via Feshbach resonances~\cite{feshbach1962feshbach,chin2010feshbach}, field gradients, and single-site resolution capabilities offered by state-of-the-art quantum gas microscopes ~\cite{thywissen2015microscope,zwierlein2016microscope} as powerful tools for qubit manipulation.

Our implementation makes use of the intra-band contact interactions as well as inter-band exchange interactions to implement logic qubit rotations and entanglement generation. We do not require continuous laser driving, nor any superlattice or spin-dependent lattice configurations. Through the additional use of a field gradient, we are able to tune the nearest-neighbour interactions between qubits to generate a desired Hamiltonian. As a proof-of-principle concept of measurement-based quantum computation~\cite{briegel2009measurementBasedQC,raussendorf2003measurementBasedQC} we show how to realize an Ising model with significantly stronger spin couplings compared to conventional superexchange, which we show can create high fidelity 1D cluster states. One can also tune the system to emulate XXZ or Heisenberg models, which may be used (as well as the Ising) to measure complex time-dependent quantities such as out-of-time-ordered correlation functions (OTOCs)~\cite{shenker2014OTOC,maldacena2016OTOC}. The interactions can be manipulated via Feshbach resonances, allowing us to turn them off and on as needed. The single-site resolution also permits the manipulation of individual atoms before and during such computations, allowing for effective error detection and insight on what the quantum system is doing on a per-site basis.

In Section II we give an overview of our model, introduce the decoherence-free subspace used to define our logic qubits, and then derive the superexchange interactions between them. In Section III we detail how these interactions can be tuned through an external field gradient, and show how to use them for cluster state generation and OTOC measurement. In Section IV we describe different protocols for qubit initialization and readout.
\section{Decoherence-free subspace qubits}
\subsection{Multi-band Fermi-Hubbard model}

\begin{figure*}[htb]
\centering
\includegraphics[width=1
\linewidth]{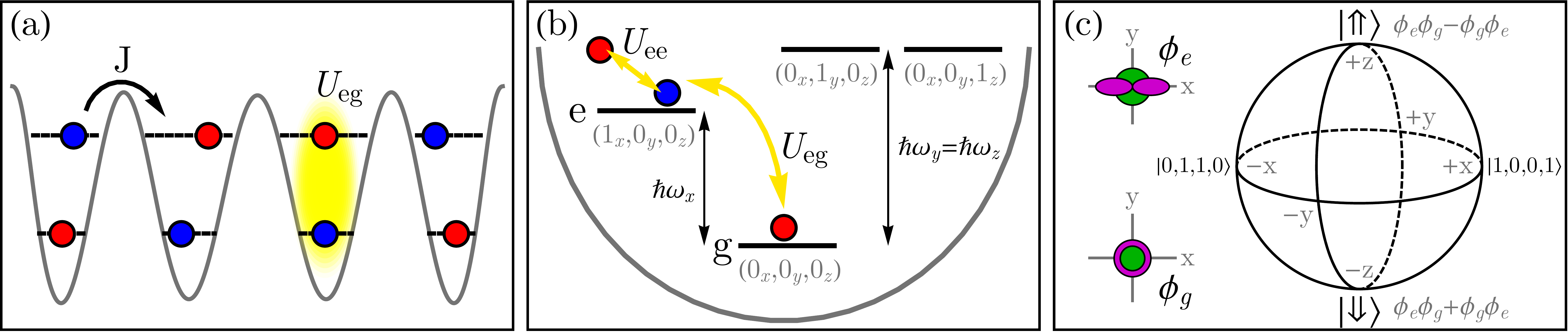}
\caption{(a) Schematic of the optical lattice setup. Atoms can only tunnel along the $\hat{x}$ direction (effective 1D system). Each site holds two atoms (one per band), which exhibit intra-band repulsion and inter-band exchange interactions. (b) Band schematic for a single lattice site. The atoms are loaded into the ground motional band $g$, and the first-excited band $e$ with one motional excitation along the $\hat{x}$ direction. Other singly-excited bands along $\hat{y}$, $\hat{z}$ are not populated, and are nondegenerate with $e$ due to higher lattice depths along $\hat{y}$, $\hat{z}$. Gray numbers in brackets are the harmonic excitation numbers, and $\omega_{\nu}$ for $\nu \in \{x,y,z\}$ are the onsite trapping frequencies. (c) Effective Bloch sphere for the decoherence-free subspace. Logical 0(1) states $\ket{\Downarrow}$($\ket{\Uparrow}$) are (anti-)symmetric superpositions of two different Wannier functions, with the opposite symmetry for the spin wavefunctions due to the overall fermionic nature of the atoms. Both these states are eigenstates of the interaction $\hat{H}_{U}$ and gradient $\hat{H}_{B}$, and are immune to uniform external magnetic fields.}
\label{fig_Schematic}
\end{figure*}

The system we work with is a 3D optical lattice populated by fermionic atoms, as depicted in Fig.~\ref{fig_Schematic}(a). The atoms are prepared in their ground electronic state, and restricted to two populated hyperfine states which we denote as $\sigma \in\{\uparrow,\downarrow\}$ acting as a spin-1/2 degree of freedom. We assume the atoms are loaded into two motional bands $e$ and $g$, with $e$ an excited band holding one motional excitation along the $\hat{x}$ direction in the harmonic approximation [as shown in Fig.~\ref{fig_Schematic}(b)], and $g$ the lowest band. The lattice depth is made more shallow along the $\hat{x}$ direction, confining the system to an effective 1D configuration and ensuring that to good approximation only the $e$ band atoms can tunnel because of the more delocalized nature of excited motional states. We assume a lattice with $L$ lattice sites along the $\hat{x}$ direction, each one populated by two atoms ($N$ atoms, $N=2L$), one in $e$ and one in $g$. The full Hamiltonian describing the system is a two-band Fermi-Hubbard model given by,
\begin{equation}
\label{eq_FermiHubbard}
    \hat{H}=\hat{H}_{J}+\hat{H}_{U}+\hat{H}_{B}.
\end{equation}
where the tunneling Hamiltonian is
\begin{equation}
    \hat{H}_{J}=-J\sum_{\langle i,j\rangle,\sigma}\left(\hat{c}_{i,e,\sigma}^{\dagger}\hat{c}_{j,e,\sigma}+h.c.\right),
\end{equation}
with $\hat{c}_{j,\mu,\sigma}$ annihilating an atom on site $j$ in band $\mu \in\{e,g\}$ with spin $\sigma$. The nearest-neighbour tunneling integral is $J$, and the lattice indexing is along $\hat{x}$ only. The atoms also have an onsite interaction, whose Hamiltonian is
\begin{equation}
\begin{aligned}
    \hat{H}_{U}&=\sum_{j}\left(U_{ee}\hat{n}_{j,e,\uparrow}\hat{n}_{j,e,\downarrow}+U_{gg}\hat{n}_{j,g,\uparrow}\hat{n}_{j,g,\downarrow}\right)\\
    &+\frac{U_{eg}}{2}\sum_{j}\left(\hat{n}_{j,e,\uparrow}\hat{n}_{j,g,\downarrow}+\hat{n}_{j,e,\downarrow}\hat{n}_{j,g,\uparrow}\right)\\
    &-\frac{U_{eg}}{2}\sum_{j}\left(\hat{c}_{j,e,\uparrow}^{\dagger}\hat{c}_{j,e,\downarrow}\hat{c}_{j,g,\downarrow}^{\dagger}\hat{c}_{j,g,\uparrow}+h.c.\right),
\end{aligned}
\end{equation}
where $\hat{n}_{j,\mu,\sigma}=\hat{c}_{j,\mu,\sigma}^{\dagger}\hat{c}_{j,\mu,\sigma}$. The first line is the interaction energy between atoms in the same band ($U_{ee}$ and $U_{gg}$ for $e$ and $g$ bands respectively), while the last two lines are direct and exchange interactions of strength $U_{eg}$ between atoms in two different bands. The values for these interactions depend on the lattice depths, but their magnitudes can be globally tuned by using a Feshbach resonance to modify the scattering length. We will operate in the strongly-interacting regime where $U_{ee}, U_{gg}, U_{eg} \gg J$. See Supplementary Material A for details on the derivation of Eq.~\eqref{eq_FermiHubbard} and its parameters.

In addition to the core lattice dynamics, we also permit an externally-imposed linear field gradient along the $\hat{x}$ direction,
\begin{equation}
    \hat{H}_{B}=\frac{B}{2}\sum_{j}j\times\left(
\hat{n}_{j,e,\uparrow}+\hat{n}_{j,g,\uparrow}-\hat{n}_{j,e,\downarrow}-\hat{n}_{j,g,\downarrow}\right),
\end{equation}
where $B$ is the energy shift between sites. This shift could be implemented with a direct magnetic field gradient, or with the synthetic magnetic gradient of a vector light shift that creates a differential potential between the two spin states. This gradient can modify the effective superexchange interactions between adjacent sites, providing a tunable knob to manipulate the system dynamics.

\subsection{Decoherence-free subspace}
The spin states $\sigma$ feel different Zeeman shifts from external magnetic fields and are therefore vulnerable to uniform magnetic field fluctuations which severely reduce coherence times. To mitigate this effect, we use equal-weight superpositions of the spin states on every lattice site which will feel no linear shift from the external field, realizing a decoherence-free subspace~\cite{lidar1998decoherenceFree}. These superpositions are the singlet and triplet states, $(\ket{\uparrow,\downarrow}\pm \ket{\downarrow,\uparrow})/\sqrt{2}$ (written in first-quantized notation for two atoms on the same site). Their corresponding spatial wavefunctions must uphold the opposite symmetries to maintain overall fermionic parity, which can be accomplished by putting them in respective symmetric and anti-symmetric superpositions of the $e$, $g$ band states, $(\ket{e,g}\mp \ket{g,e})/\sqrt{2}$. The states for a single lattice site may thus be written as,
\begin{equation}
\begin{aligned}
\ket{\Uparrow}&=\frac{1}{2}\left(\ket{\uparrow,\downarrow}+\ket{\downarrow,\uparrow}\right)\left(\ket{e,g}-\ket{g,e}\right)\\
&=\frac{1}{\sqrt{2}}\left(\ket{1,0,0,1}+\ket{0,1,1,0}\right),\\
\ket{\Downarrow}&=\frac{1}{2}\left(\ket{\uparrow,\downarrow}-\ket{\downarrow,\uparrow}\right)\left(\ket{e,g}+\ket{g,e}\right)\\
&=\frac{1}{\sqrt{2}}\left(\ket{1,0,0,1}-\ket{0,1,1,0}\right),
\end{aligned}
\end{equation}
where in the second lines of each equation we have rewritten the states in second quantization assuming a Fock basis ordering of $\ket{n_{e,\uparrow},n_{e,\downarrow},n_{g,\uparrow},n_{g,\downarrow}}$. These states are also eigenstates of the interaction Hamiltonian,
\begin{equation}
\begin{aligned}
\hat{H}_{U} \ket{\Uparrow}&=0,\\
\hat{H}_{U} \ket{\Downarrow}&=U_{eg}\ket{\Downarrow},
\end{aligned}
\end{equation}
and zero-energy eigenstates of the Zeeman Hamiltonian, $\hat{H}_{B}\ket{\Uparrow}=\hat{H}_{B}\ket{\Downarrow}=0$. For experimentally realistic parameters (see Supplementary Material A), the exchange interaction energy difference $U_{eg}$ between these states can exhibit a $> 100$-fold reduction in sensitivity to external magnetic field fluctuations compared to the shifts that bare nuclear-spin states $\{e,g\}$ would experience.

Altogether, the decoherence-free states $\{\ket{\Uparrow}$, $\ket{\Downarrow}\}$ form a spin-1/2 logical qubit subspace on each site of the optical lattice, as shown in Fig.~\ref{fig_Schematic}(c). We can have states along different axes of the associated Bloch sphere such as $\ket{\Rightarrow} = (\ket{\Uparrow}+\ket{\Downarrow})/\sqrt{2} = \ket{1,0,0,1}$ and $\ket{\Leftarrow} = (\ket{\Uparrow}-\ket{\Downarrow})/\sqrt{2}=\ket{0,1,1,0}$, which are maximally entangled between the spin and motional degrees of freedom. Onsite qubit rotations can be made without leaving the subspace as discussed in Section IV. We are also robust to unwanted band-changing collisions into other singly-excited bands (see Supplementary Material E for benchmarking). Nearest-neighbour tunneling processes can be used to generate superexchange interactions and create entanglement between the qubits, as we will describe next.

\begin{figure}
\centering
\includegraphics[width=1\linewidth]{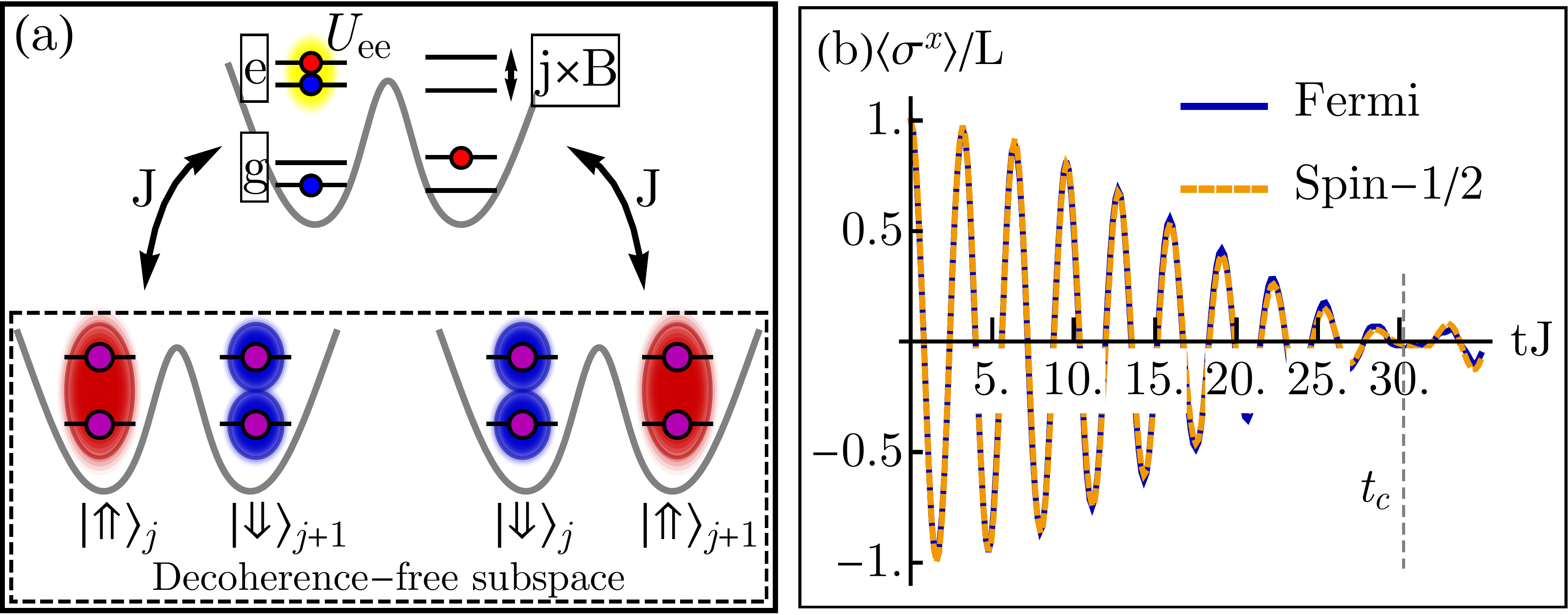}
\caption{(a) Schematic for superexchange in the multi-band system. States in the decoherence-free subspace are coupled through a second-order virtual tunneling process by atoms in the $e$ band. The intermediate state will have two atoms in the same band on the same site, exhibiting $U_{ee}$ repulsion. Provided the energy gaps between the occupied and intermediate states are large compared to $J$, this leads to a spin-like interaction in the decoherence-free subspace. (b) Benchmark comparison between the full Fermi-Hubbard model of Eq.~\ref{eq_FermiHubbard} and the superexchange model of Eq.~\eqref{eq_SpinModel}. Parameters are $U_{ee}/J = 179$, $U_{eg}/J=239$ and $B/J=369$. The time $t_c$ is the point where $\langle \hat{\sigma}^{x}\rangle$ vanishes, corresponding to the time needed to make a cluster state (see Section III). Note that the oscillation frequency seen is not representative, and is an artifact of finite point sampling; the genuine oscillation frequency is set by $U_{eg}$.}
\label{fig_Superexchange}
\end{figure}

\subsection{Superexchange Hamiltonian}
The single-site states in our decoherence-free subspace have energies of $0$ or $U_{eg}$ from the exchange interaction. However, if an $e$-band atom tries to tunnel into an adjacent site, the resulting state will feel an intra-band $U_{ee}$ interaction, as depicted in Fig.~\ref{fig_Superexchange}(a) ($U_{gg}$ plays no role as $g$ atoms do not tunnel). So long as the energy difference between the configurations before and after tunneling is much larger than the tunneling amplitude $J$, no resonant tunneling will occur. However, the system will exhibit a second-order interaction through virtual processes starting and ending in the decoherence-free subspace, with an offresonant excited state in-between. This leads to an effective superexchange Hamiltonian $\hat{H}_{ex}$ between neighbouring lattice sites~\cite{trotzky2008superexchange}~\cite{bravyi2011schriefferWolff} with  matrix elements between states $\ket{i}$, $\ket{j}$ in the decoherence-free subspace given by
\begin{equation}
\begin{aligned}
\label{eq_Superexchange}
    \bra{i}\hat{H}_{\mathrm{ex}}\ket{j} &= \sum_{k}\Delta_{ijk}\bra{i}\hat{H}_{J}\ket{k}\bra{k}\hat{H}_{J}\ket{j},\\
    \Delta_{ijk}&=\frac{1}{2}\left(\frac{1}{E_{i}-E_{k}}+\frac{1}{E_{j}-E_{k}}\right),
\end{aligned}
\end{equation}
where $E_{i}$ is the unperturbed state energy $(\hat{H}_{U}+\hat{H}_{B})\ket{i}=E_{i}\ket{i}$, and $k$ runs over all excited states outside the decoherence-free subspace (see Supplementary Material C for details). Note that we must also keep the unperturbed $\hat{H}_{U}+\hat{H}_{B}$ in the dynamics.

After some algebra, we find that the superexchange interaction takes the form of
\begin{widetext}
\begin{equation}
\begin{aligned}
\label{eq_SpinModel}
\hat{H}_{\mathrm{ex}}&=J_{\parallel}\sum_{\langle i,j\rangle}\hat{\sigma}_{i}^{x}\hat{\sigma}_{j}^{x}+J_{\perp}\sum_{\langle i,j\rangle}\left(\hat{\sigma}_{i}^{y}\hat{\sigma}_{j}^{y}+\hat{\sigma}_{i}^{z}\hat{\sigma}_{j}^{z}\right)+J_{xz}\sum_{\langle i,j\rangle}\left(\hat{\sigma}_{i}^{x}\hat{\sigma}_{j}^{z}-\hat{\sigma}_{i}^{z}\hat{\sigma}_{j}^{x}\right)+J_{z}\sum_{j}\hat{\sigma}_{j}^{z}\\
&\approx \sum_{\langle i,j\rangle}\left[\frac{J_{\parallel}+J_{\perp}}{2}\left(\hat{\sigma}_{i}^{x}\hat{\sigma}_{j}^{x}+\hat{\sigma}_{i}^{y}\hat{\sigma}_{j}^{y}\right)+J_{\perp}\hat{\sigma}_{i}^{z}\hat{\sigma}_{j}^{z}\right]+J_{z}\sum_{j}\hat{\sigma}_{j}^{z}.
\end{aligned}
\end{equation}
\end{widetext}
Here, $\hat{\sigma}_{j}^{\nu}$ are standard Pauli operators acting on the $\{\ket{\Uparrow},\ket{\Downarrow}\}$ logic qubits at site $j$. The coefficients depend on three particular combinations of the repulsive and exchange interactions that show up in the energy denominators of Eq.~\ref{eq_Superexchange},
\begin{equation}
\begin{aligned}
\label{eq_UGamma}
U_1 &=2U_{ee}+U_{eg},\\
U_2 &=2U_{ee}-U_{eg},\\
U_3 &=2U_{ee}-3U_{eg},
\end{aligned}
\end{equation}
The corresponding coupling constants are
\begin{equation}
\begin{aligned}
J_{\parallel}&=-\frac{J^2}{2}\left(\frac{U_1}{B^2-U_1^2}+\frac{2U_2}{B^2-U_2^2}+\frac{U_3}{B^2-U_3^2}\right),\\
J_{\perp}&=\frac{J^2 U_2 \left(3 B^2+U_1 U_3\right) (U_1-U_2)^2}{\left(B^2-U_1^2\right) \left(B^2-U_2^2\right) \left(B^2-U_3^2\right)},\\
J_{xz}&=-\frac{2 B J^2 U_2 (U_1-U_2)}{\left(B^2-U_1^2\right) \left(B^2-U_3^2\right)},\\
J_{z}&=J^2\left(\frac{U_1}{B^2-U_{1}^2}-\frac{U_3}{B^2-U_{3}^2}\right)-\frac{U_1-U_2}{4}.
\end{aligned}
\end{equation}

Note that in going to the second line of Eq.~\eqref{eq_SpinModel}, we have used the fact that the external field $\sim J_{z}\hat{\sigma}^{z}$ imposed by the bare exchange interactions [i.e. the $(U_1-U_2)/4$ in $J_{z}$ coming from $\hat{H}_{U}$] is much larger than any superexchange processes. As a result, some of the terms in the first line of Eq.~\eqref{eq_SpinModel} can be neglected. To good approximation, the system is projected into a given $\hat{\sigma}^{z}$-eigenvalue manifold, causing the $J_{xz}$ term to be negligible in a rotating-wave approximation. Furthermore, any $\hat{\sigma}^{x}\hat{\sigma}^{x}$ or $\hat{\sigma}^{y}\hat{\sigma}^{y}$ terms will be projected into a $(\hat{\sigma}^{+}\hat{\sigma}^{-}+h.c.)$ form. This allows us to recollect them and write the Hamiltonian as an effective XXZ-type model with an external field.

This model is valid so long as all of the denominators are large compared to the tunneling, i.e. $|B-U_{\gamma}| \gg J$ for $\gamma \in \{1,2,3\}$, to avoid higher-order effects. Fig.~\ref{fig_Superexchange}(b) shows a comparison between the full Fermi-Hubbard dynamics and the superexchange model for parameters of interest, looking at collective observable $\langle \hat{\sigma}^{x}\rangle = \sum_{j}\langle \hat{\sigma}^{x}_{j}\rangle$ (section IV discusses its measurement).

\section{Entanglement generation}
\subsection{Interaction form}

While the form of the interaction coefficients in our superexchange model is nontrivial, the key features are the non-identical denominators. By tuning the parameters such that some of the denominators become much smaller than the others, we can choose which interactions get turned on and off. This permits the isolation of terms of interest, which can then be employed for useful entanglement generation.

\begin{figure}[h]
\centering
\includegraphics[width=1
\linewidth]{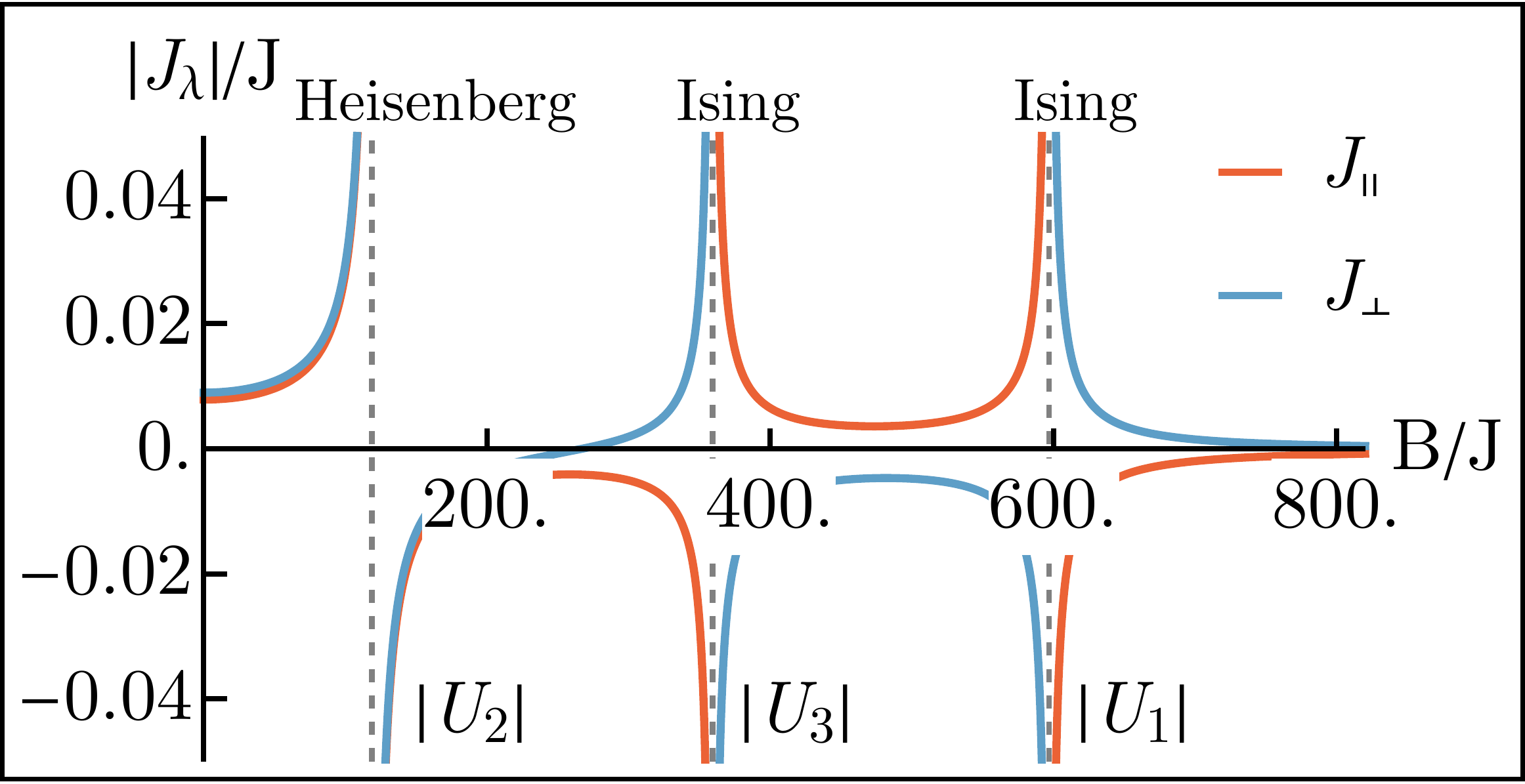}
\caption{Field gradient dependence of the superexchange interaction parameters $J_{\parallel}$ and $J_{\perp}$. The interaction strengths are $U_{ee}/J = 179$ and $U_{eg}=239$, leading to resonant denominator values of $U_{1}/J=597$, $U_{2}/J=119$, $U_{3}/J=-359$. All of the resonance points where $B=|U_{\gamma}|$ for $\gamma \in \{1,2,3\}$ are marked with dashed lines.}
\label{fig_BField}
\end{figure}

Fig.~\ref{fig_BField} plots $J_{\parallel}$ and $J_{\perp}$ for relevant lattice depths as a function of field gradient $B$. We observe three resonances, where $B = U_{\gamma}$ for some $\gamma$ and the corresponding denominator vanishes. While our second-order perturbative Hamiltonian will not be correct at those resonance points due to higher-order effects, if we stay close to them while still obeying $|B-|U_{\gamma}||\gg J$, the superexchange Hamiltonian will remain valid and the near-resonant terms will dominate the dynamics.

Operating close to the resonance conditions $B \approx |U_{3}|$ or $B \approx |U_{1}|$ is of particular utility since in this case, $J_{\parallel} \approx - J_{\perp}$ and to leading order the $\hat{\sigma}^{x}\hat{\sigma}^{x}$ and $\hat{\sigma}^{y}\hat{\sigma}^{y}$ terms are cancelled out. We are left with just an Ising model,
\begin{equation}
\label{eq_Ising}
    \lim_{B \to |U_{3}|, |U_{1}|} \hat{H}_{\mathrm{ex}} = J_{zz}\sum_{\langle i,j\rangle} \hat{\sigma}_{i}^{z}\hat{\sigma}_{j}^{z}+J_{z}\sum_{j}\hat{\sigma}^{z},
\end{equation}
where $J_{zz} \approx J_{\perp} \approx - J_{\parallel}$. Since we cannot be exactly at resonance, there will be some discrepancy between the two coefficients; a sufficient approximation is to take the average, $J_{zz} \approx (J_{\perp}-J_{\parallel})/2$. The nearest-neighbour Ising model has a wide variety of applications for entanglement generation between qubits, and can be used to make cluster states as demonstrated in the next section. Furthermore, the Ising interaction strength near the resonance scales as $\sim J^2 / |B - |U_{\gamma}||$, which is much faster than conventional superexchange $\sim J^2/U_{\gamma}$ (the former's denominator is on the order of $\sim 10J$, while the latter is $\sim 100J$).

The system's tunability also permits realization of other spin models. If we instead bring the field gradient close to the $U_{2}$ resonance, the interaction coefficients are equal in magnitude and sign, which creates a Heisenberg model with a transverse field:
\begin{equation}
    \lim_{B \to |U_{2}|} \hat{H}_{\mathrm{ex}} = J_{\perp}\sum_{\langle i,j\rangle} \vec{\sigma}_{i}\vec{\sigma}_{j}+J_{z}\sum_{j}\hat{\sigma}^{z}.
\end{equation}
Finally, as seen from Eq.~\eqref{eq_SpinModel}, we can realize the XXZ model with a wide range of coefficients, which has seen recent experimental interest~\cite{dimitrova2019tiltedSuperexchange}.

The homogeneity (how close $|J_{\parallel}|$ can be made to $|J_{\perp}|$) and positions of the resonance points depend on the interaction strength. Having higher $U_{ee}/J$, $U_{eg}/J$ leads to better homogeneity, but at the cost of moving the resonance points outwards and thus requiring a larger field gradient $B/J$. This requirement can be mitigated by reducing the tunneling rate $J$, implying a compromise between gradient strength and experimental timescale. Increasing $U_{ee}$, $U_{eg}$ without affecting $J$ is done by either increasing transverse lattice depth along the $\hat{y}$, $\hat{z}$ directions or increasing the scattering length $a_s$ via Feshbach resonance. See Supplementary Material A for a more detailed discussion of experimental parameters used throughout this text. 

\subsection{Cluster state generation}
The Ising model in Eq.~\eqref{eq_Ising} can be employed to generate useful entanglement for quantum computation. We can generate a cluster state, which is a multipartite entangled state used for measurement-based quantum computing~\cite{briegel2009measurementBasedQC}. A cluster state is a resource that can reproduce the results of circuit-based computations without needing explicit entangling gates between individual qubits. All entanglement generation is front-loaded into the cluster state itself. Once this state is prepared, a computation is done by successive feed-forward measurements. Given the long coherence times and innate 3D nature of the lattice, we can not only generate a single cluster state, but an entire array of them to be used or sorted as needed. While a 1D cluster state alone is not sufficient for universal computation, our protocol can be extended to 2D in a straightforward manner as discussed in Section V. We focus on 1D as a proof-of-principle demonstration of the system's capabilities.

A cluster state is defined as,
\begin{equation}
\label{eq_ClusterStateDefinition}
\ket{\psi_{c}}=\prod_{\langle i,j\rangle}\exp \left[-i\frac{\pi}{4}\left( \hat{\sigma}_{i}^{z}\hat{\sigma}_{j}^{z}-\hat{\sigma}_{i}^{z}-\hat{\sigma}_{j}^{z}\right)\right]\ket{\psi(0)},
\end{equation}
where the exponential operator is a controlled phase gate, applied across all possible nearest-neighbour links. The initial-state corresponds to a spin-polarized state along $+\hat{x}$,
\begin{equation}
\label{eq_InitialCondition}
\ket{\psi(0)}=\prod_{j}\ket{\Rightarrow}_{j}.
\end{equation}
As seen from Eq.~\eqref{eq_Ising}, our system already contains the necessary Ising interaction. The protocol we use is similar to the one discussed in Ref.~\cite{mamaev2019cluster}, but without the need for continuous laser driving and easier to tune with the field gradient. We simply prepare the initial state $\ket{\psi(0)}$ (as described in section IV), quench the field gradient or interaction strength (via Feshbach resonance) to satisfy $B \approx |U_3|$, and wait for a time
\begin{equation}
t_c = \pi/(4J_{zz}),\>\>\>\>J_{zz}\approx(J_{\perp}-J_{\parallel})/2.
\end{equation}
This will implement the cross-site terms in the controlled phase gate. The only remaining task is to implement the single-particle terms. Since our $J_{z}$ is so much larger than the interaction and commutes with it (to good approximation), we could in principle determine how many full periods of single-particle evolution have occurred during $t_c$, and then compensate by letting the system evolve further such that the total time spent is $\pi/(2J_{z}) \text{ mod } \pi/{J_{z}}$. However, a more prudent approach is to use a spin-echo $\pi$ pulse $\hat{\Pi}=e^{-i \pi \hat{\sigma}^{x}/2}$ halfway through the evolution. This will cause any phase accrued from the single-particle terms to be undone by itself during the second half of the evolution. In addition, such a pulse can help with unwanted sources of noise that are not captured by our model. After this evolve-echo-evolve sequence is done, we let the system evolve further for a time $\pi/(2J_{z})$ for the necessary single-particle rotation, and then stop the dynamics by turning off the tunneling and interactions (via Feshbach resonance). The overall protocol is thus,
\begin{equation}
\label{eq_ClusterProtocol}
    \ket{\psi_c} = e^{-i \hat{H}_{ex} \pi/(2 J_z)}e^{-i \hat{H}_{ex} t_c /2} \hat{\Pi} e^{-i \hat{H}_{ex} t_c /2}\ket{\psi(0)}.
\end{equation}

\begin{figure}
\centering
\includegraphics[width=1
\linewidth]{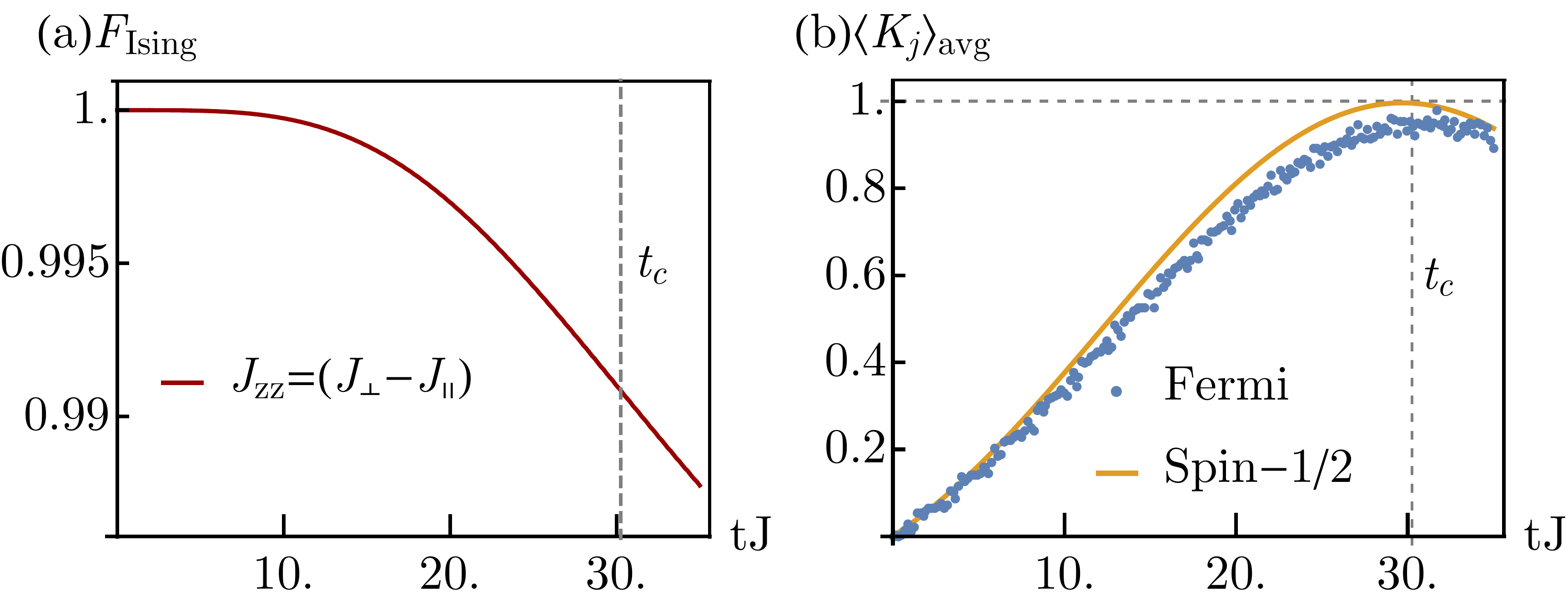}
\caption{(a) Fidelity between the state generated by the spin-echo protocol of Eq.~\eqref{eq_ClusterProtocol}, and the state generated by an ideal Ising model with $J_{zz} = (J_{\perp}-J_{\parallel})/2$. Parameters are $U_1/J = 597$, $U_2/J = 119$, $U_3/J = -359$ and $B/J = 369$. System size is $L=8$. (b) Stabilizer values $\langle \hat{K}\rangle_{j}$ for the spin-echo protocol using the superexchange model $\hat{H}_{ex}$ (orange line) and full Fermi-Hubbard model $\hat{H}$ (blue dots), for system size $L=4$. The stabilizer operators in the corners are defined differently: The missing $\hat{\sigma}^{z}_{j}$ is absent from the correlator, and an overall rotation of $e^{-i \pi \hat{\sigma}^{z}/4}$ is applied to the state before calculating them, since they are missing such a rotation from the vacant link on the open end.}
\label{fig_Cluster}
\end{figure}

Fig.~\ref{fig_Cluster}(a) shows state fidelity between time-evolution using the echo protocol with our superexchange model of Eq.~\eqref{eq_SpinModel} and an ideal Ising model evolution which can generate a perfect cluster state. Inhomogeneity between $J_{\parallel}$ and $J_{\perp}$ and the breakdown of the rotating-wave approximation (since $J_{z}$ is not infinite) are the main sources of error, but in general they can be made small.

One of the core advantages to measurement-based quantum computation is that the quality of a cluster state can be estimated using spatially local properties rather than global fidelity. To this end, we look at multi-body correlators called stabilizer operators, defined as
\begin{equation}
\hat{K}_{j} = \hat{\sigma}_{j}^{x}\prod_{\langle i,j\rangle}\hat{\sigma}_{i}^{z},
\end{equation}
i.e. products of an $\hat{\sigma}^{x}$ measurement on one site and $\hat{\sigma}^{z}$ on all its neighbouring sites (thus a 3-body operator in 1D, 5 in 2D and 7 in 3D). An ideal cluster state is an eigenstate of all such stabilizer operators with eigenvalue $1$,
\begin{equation}
\hat{K}_{j}\ket{\psi_c} = +\ket{\psi_c},\>\>\>\>\bra{\psi_c}\hat{K}_{j}\ket{\psi_c} = +1.
\end{equation}
Note that the sign of the eigenvalue does not matter as long as it is the same for all sites, since we can flip it by applying a collective $\hat{\sigma}^{z}$ rotation. The closeness of each stabilizer correlator $\langle \hat{K}_{j}\rangle$ to $1$ serves as a local measure of cluster state quality. If we average it over a region of the lattice, $\langle \hat{K}_{j}\rangle_{\mathrm{avg}}=\sum_{j}\langle \hat{K}_{j}\rangle$, we get an estimate of the fidelity for computations done using that region.

Fig.~\ref{fig_Cluster}(b) shows the average value of these cluster correlators for a small system as our protocol goes on. Note that in our simulations the lattice has boundaries, which means that the corner sites have their respective stabilizer operators defined differently (see caption). The main deviations come from inhomogeneity between $J_{\parallel}$ and $J_{\perp}$, as well as leakage of population into the excited states due to higher-order processes. These can be remedied by driving further from resonance (at the cost of longer $t_c$), and using larger interactions to keep the coefficients homogeneous (at the cost of needing a stronger gradient). However, we still find high stabilizer values $\langle \hat{K}_{j}\rangle_{\text{avg}} > 0.95$ indicating a good cluster state.

\subsection{OTOC measurement}

Another powerful feature of the tunable superexchange coefficients (c.f. Fig.~\ref{fig_BField}) is their ability to change sign. We can quench the gradient from just below one of the resonances, say $B \lesssim |U_3|$, to just above, $B \gtrsim |U_3|$. This flips the interaction coefficient, allowing for the implementation of a unitary time-reversal. Such capabilities are applicable to the study of out-of-time ordered correlation functions (OTOCs), which have garnered much recent interest due to their applications in quantum chaos, butterfly effects and temporal correlation spreading~\cite{maldacena2016OTOC}.

An OTOC is a two-time, four-operator correlator defined as $\langle \hat{W}^{\dagger}(t) \hat{V} \hat{W}(t) \hat{V}\rangle$, where $\hat{W}$, $\hat{V}$ are time-independent commuting operators and $\hat{W}(t) = e^{i \hat{H} t} \hat{W} e^{-i \hat{H} t}$. The connection to chaos can be understood by considering $\hat{W}$, $\hat{V}$ to be local operators with some spatial separation. When the system is initialized, they commute and the corresponding OTOC is zero. As correlations spread, the spatial extent of $\hat{W}$ and thus the OTOC increase. The slope of increase gives information about the propagation of correlations in the system.

\begin{figure}
\centering
\includegraphics[width=1\linewidth]{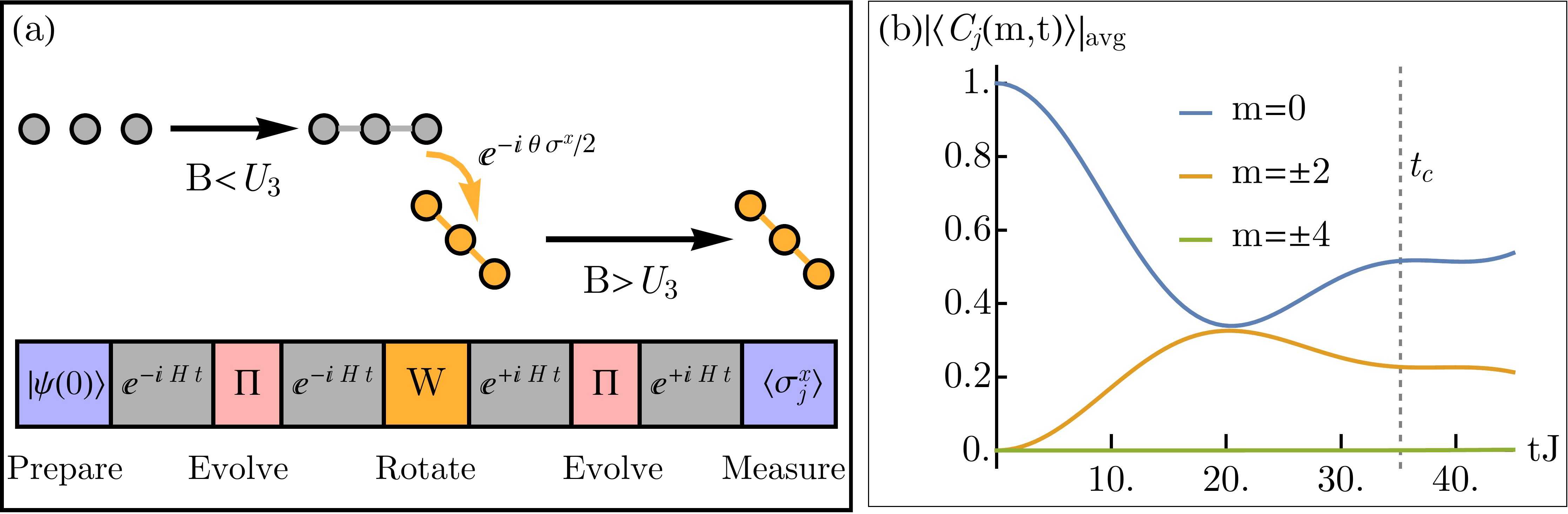}
\caption{(a) Schematic of the OTOC-measuring protocol of Eq.~\eqref{eq_OTOCProtocol}. The sign of the Hamiltonian is changed by quenching the field gradient $B$ from below to above a resonance (here the $U_3$ resonance), choosing the $J_{\perp}$ values to have equal magnitude and opposite sign. $\pi$ pulses are done in the middle of each evolution to remove unwanted single-particle rotations. (b) Fourier-transformed OTOC components [c.f. Eq.~\eqref{eq_OTOCFourierTransform}], averaged over all $j$ and computed with the superexchange model. System size is L = 8 (open boundaries). Parameters are $U_1/J = 597$, $U_2/J = 119$, $U_3/J = -359$, with the magnetic field quenched from $B/J = 350$ to $\tilde{B}/J = 370$ about the $U_3$ resonance, corresponding to $J_{\perp}/J = \pm 0.025$.}
\label{fig_OTOC}
\end{figure}

In our case, we can measure an OTOC in a straightforward manner. We time-evolve under the effective superexchange Hamiltonian to a time $t$, apply a spin rotation by some angle $\theta$, quench the field gradient to flip the interaction sign, evolve for another $t$, and finally measure some observable (total evolution time $2t$), similar to the sequence used in Ref.~\cite{garttner2017OTOC}. If we start in $\ket{\psi(0)}$, a sample such protocol may be written as,
\begin{widetext}
\begin{equation}
\label{eq_OTOCProtocol}
\begin{aligned}
\langle \hat{C}_{j}(\theta,t)\rangle &= \bra{\psi(0)}\left(e^{+i \hat{H}_{\mathrm{ex}}t} e^{-i \frac{\theta}{2} \hat{\sigma}^{x}} e^{-i \hat{H}_{\mathrm{ex}}t}\right) \hat{\sigma}_{j}^{x} \left(e^{+i \hat{H}_{\mathrm{ex}}t} e^{-i \frac{\theta}{2} \hat{\sigma}^{x}} e^{-i \hat{H}_{\mathrm{ex}}t}\right)\ket{\psi(0)}\\
&= \bra{\psi(0)} \hat{W}^{\dagger}(t) \hat{V} \hat{W}(t) \hat{V}\ket{\psi(0)},
\end{aligned}
\end{equation}
\end{widetext}
where $\hat{W} = e^{-i \frac{\theta}{2} \hat{\sigma}^{x}}$, and $\hat{V} = \hat{\sigma}_{j}^{x}$. In going to the second line, we have used the fact that $\hat{\sigma}_{j}^{x}\ket{\psi(0)}=\ket{\psi(0)}$. Fig.~\ref{fig_OTOC}(a) depicts this protocol.

The sign change can be done about any $U_{\gamma}$ resonance (c.f. Fig.~\ref{fig_BField}). Here we will use the Ising model with $U_3$. Note that each half of the time-evolution will also include a spin-echo midway to remove the $\hat{\sigma}^{z}$ single-particle rotations, since their sign cannot be fully reversed. The total implementation is thus,
\begin{equation}
\begin{aligned}
e^{-i \hat{H}_{\mathrm{ex}}t} = e^{-i \hat{H}_{\mathrm{ex}}(B)t/2} \hat{\Pi} e^{-i \hat{H}_{\mathrm{ex}}(B)t/2},\\
e^{+i \hat{H}_{\mathrm{ex}}t} = e^{-i \hat{H}_{\mathrm{ex}}(B')t/2} \hat{\Pi} e^{-i \hat{H}_{\mathrm{ex}}(B')t/2},
\end{aligned}
\end{equation}
with $B$, $B'$ on opposite sides of the chosen resonance, set such that the magnitude of $J_{zz}$ is equal for both. For the parameters of $U_1/J = 597$, $U_2/J = 119$, $U_3/J = -359$, we can quench the gradient about the $U_{3}$ resonance from $B/J = 350$ to $B'/J = 370$, yielding effective interaction strengths of $J_{\perp}/J \approx \pm 0.025$.

We can measure different OTOCs depending on evolution time $t$ and rotation angle $\theta$. For $\theta = 0$, we have trivial unitary reversal, and the corresponding OTOC would be equal to unity. As the angle increases, the OTOC exhibits a decaying slope as correlations build up. Another useful piece of information that can be extracted is the spectral fourier transform of the OTOCs,
\begin{equation}
\label{eq_OTOCFourierTransform}
\langle \hat{C}_{j}(m,t)\rangle = \sum_{\theta} e^{i m \theta}\langle \hat{C}_{j}(\theta,t)\rangle,
\end{equation}
where $\theta = \frac{2\pi n}{L}$ and $n,m \in \{0, \dots, L-1\}$. The different $m$ components give information about the system's connectivity~\cite{garttner2017OTOC}. In a 1D lattice with two neighbours per site, we should see signals at $m = \pm 2$. A 2D lattice would have signals at $\pm 4,\pm 2$, and a 3D lattice at $\pm 6, \pm 4, \pm 2$. Fig.~\ref{fig_OTOC}(b) shows these fourier-transformed OTOC values for our superexchange model. Note that we do not see any special peak or trough at $t = t_c$ because our system has open boundaries, and thus the OTOCs will feel edge effects since we average over every site. However, site-resolved OTOCs can be probed in this system with a quantum gas microscope.

\section{Preparation and control tools}
\subsection{State preparation}
In this section, we detail an explicit experimental protocol for preparing atoms in the required decoherence-free subspace. The starting point is a band insulator (two atoms of opposite spin per site) in the lowest band $g$, which can be prepared by standard cooling techniques~\cite{mazurenko2016cooling,hart2015cooling,boll2016cooling}. We then use a Raman scheme to selectively drive one of the atoms (the spin-$\downarrow$ one) into the $e$ band. The result is a state of the form $\ket{1,0,0,1} = \ket{\Rightarrow}$ on every site, which can immediately be used for cluster state generation. One can also rotate this state into others using the protocols of the next section.

\begin{figure}
\centering
\includegraphics[width=1\linewidth]{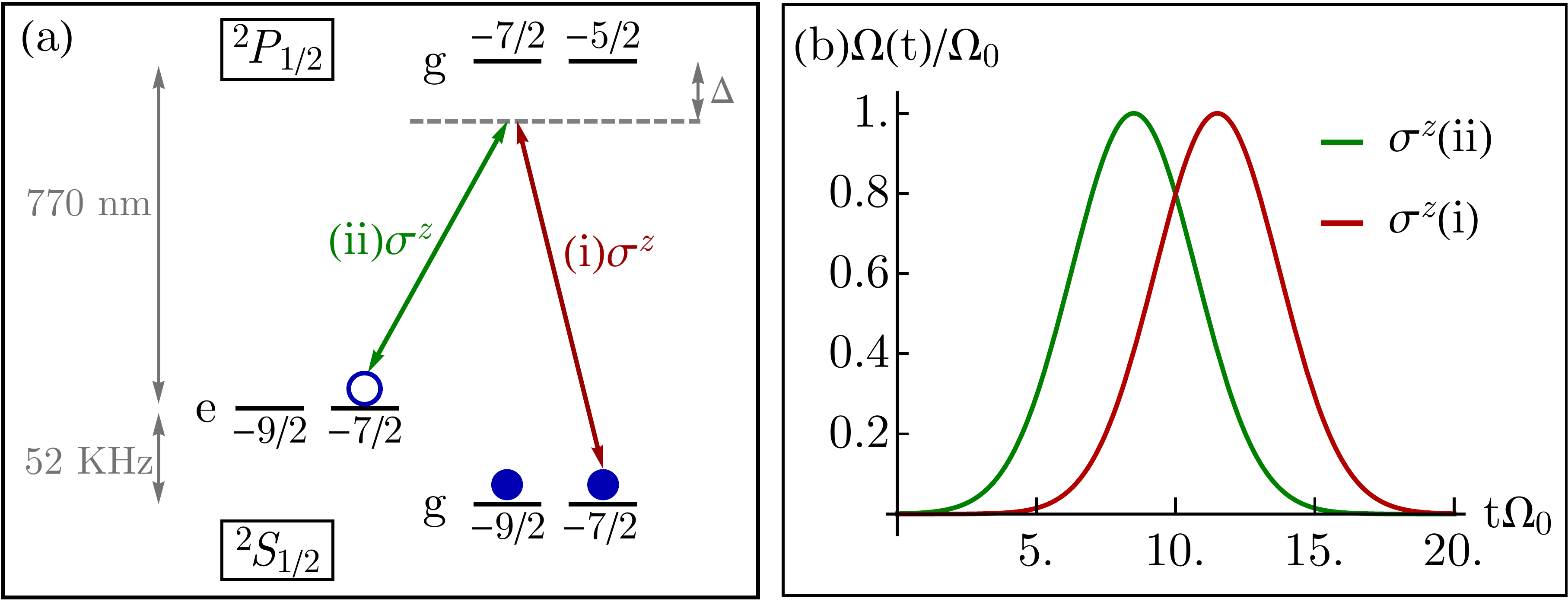}
\caption{(a) Raman scheme for transferring an $\uparrow$ atom to the $e$ band. Two linearly polarized lasers (i) and (ii) couple the states $\ket{^{2}S_{1/2}, g; 9/2, -7/2}$ and $\ket{^{2}S_{1/2}, e; 9/2, -7/2}$ through an excited state $\ket{^{2}P_{1/2},g;7/2,-7/2}$ (transition wavelength $\approx 770$ nm, unpopulated due to large detuning). Here the state label is $\ket{\Lambda,\mu;F,m_F}$, with electronic state $\Lambda$, band $\mu$, nuclear-spin $F$ and projection $m_F$ (see Supplementary Material for details). A $\pi$ pulse effects the coherent transfer. (b) STIRAP scheme for the same transfer. Instead of always-on constant Rabi frequencies for both lasers, we first ramp up the (ii) laser coupling the initially-unoccupied states with a Gaussian profile, then ramp it down while ramping up the (i) laser. The state is adiabatically dragged into the desired configuration. Here $\Omega_0$ is the maximum Rabi frequency the lasers reach, equal for both.}
\label{fig_StatePreparation}
\end{figure}

The process is depicted in Fig.~\ref{fig_StatePreparation}(a). We use a linearly-polarized laser pair to couple an occupied state in the $g$ band to an empty target state in the $e$ band via an intermediate excited state in a higher electronic level. The laser detunings $\Delta$ from the excited state are chosen to be equal and large compared to their Rabi frequencies, creating a Raman coupling between the occupied and target states. The other first-excited bands along $\hat{y}$, $\hat{z}$ do not participate because of their energy shift due to unequal lattice depth. Doing a $\pi$ pulse implements a coherent transfer, populating the $e$ band with one $\downarrow$ atom. See Supplementary Material B for more details.

If a coherent $\pi$ pulse with the above setup poses experimental challenges, one may instead use a STIRAP protocol~\cite{oreg1984stirap,bergmann1998stirap}. The laser configurations are the same, but instead we slowly ramp the laser intensities with offset Gaussian profiles as depicted in Fig.~\ref{fig_StatePreparation}(b). The coupling between the initially-unoccupied states [laser (ii)] is counter-intuitively ramped up first, and then the coupling between the initially-occupied and excited states [laser (i)] is ramped up while the previous coupling is reduced. The system follows an adiabatic evolution where the ground-state is transferred into a final configuration with the atom in the target state. Unlike the previous protocol, STIRAP can function even with $\Delta=0$, and enjoys better robustness to laser noise or bandwidth limitations.

\subsection{Rotations}

Quantum computation requires readily accessible rotations upon the qubits in the decoherence-free subspace. A $\hat{\sigma}^{z}$-type rotation is straightforward because the exchange interaction creates such a term implicitly. We tune $U_{eg}$ to be large enough for fast rotations and far from any resonances, then wait the desired time. Onsite rotations of this type can be implemented with a focused laser to change the lattice potential (and thus interactions) that a site feels.

A $\hat{\sigma}^{x}$ rotation may also be realized by using additional external ingredients. One method is to use an on-site field gradient. A linear gradient is no longer sufficient, because it has no on-site effect on the decoherence-free subspace (see Supplementary Material D). While a linear gradient was used to tune the cross-site superexchange interactions, there it only shifted the energies of the states involved: adjacent qubits are tunnel-coupled. For direct coupling between the decoherence-free states, a quadratic field variation is required. For example, an effective magnetic curvature could be realized via state-dependent optical potentials for the bare spin states $\sigma$ using the differential magnetic moments and vector Stark shifts. One could also apply a magnetic field curvature directly. The Hamiltonian for such a gradient on a site $j$ is,
\begin{equation}
\hat{H}_{\delta B} = \delta\Omega (x-x_0)^{2} \left(\hat{n}_{j,\uparrow}-\hat{n}_{j,\downarrow}\right).
\end{equation}
where the gradient is along the tunneling direction $\hat{x}$, focused near one lattice site and centered about some position $x_0$. Evaluating the effect of this gradient for a single site (see Supplementary Material D), we arrive at the following Hamiltonian in the basis $\{\ket{\Uparrow}, \ket{\Downarrow}\}$ (on site $j$):
\begin{equation}
\label{eq_HamiltonianDecoherenceFreeWithGradient}
\hat{H}_{\mathrm{site}} = \left(\begin{array}{cc} 0 &\frac{\hbar \>\delta\Omega}{m \omega_x} \\\frac{\hbar \>\delta\Omega}{m \omega_x}& U_{eg}\end{array}\right) =  -\frac{U_{eg}}{2}\hat{\sigma}^{z}_{j}+\frac{\hbar \>\delta\Omega}{m \omega_x} \hat{\sigma}^{x}_{j}.
\end{equation}
A $\hat{\sigma}^{x}$ rotation may be done by either turning the interactions off via Feshbach resonance, or ensuring that $\hbar\>\delta\Omega/(m\omega_x) \gg U_{eg}$. For most cases, the former will be the better choice as we do not want arbitrary on-site exchange interactions acting on the qubits unless we are generating a cluster state. One could also overcome the $U_{eg}$ interactions with a high $\delta\Omega$; while this would need a very strong field gradient, having such a requirement also shows a degree of robustness to unwanted magnetic curvature (see discussion in Supplementary Material D).

A second method of applying $\hat{\sigma}^{x}$ rotations is to use differential Stark shifts. The desired rotation operator takes the form of $\ket{1,0,0,1}\bra{1,0,0,1}-\ket{0,1,1,0}\bra{0,1,1,0}$ in the Fock basis, which can be implemented by energetically shifting one of the $\{\ket{\Rightarrow}, \ket{\Leftarrow}\}$ states in the decoherence-free subspace (i.e. applying a $\hat{\sigma}^{z}$ rotation in the $\{\ket{1,0,0,1}, \ket{0,1,1,0}\}$ basis). This may be done with a laser such as (i) in Fig.~\ref{fig_StatePreparation}(a). The detuning $\Delta$ is set sufficiently large compared to the laser Rabi frequency $\Omega$ that no population transfer occurs; the lowest-order effect is then to shift the state $\ket{1,0,0,1}$ by $\Omega^2/\Delta$, realizing the desired rotation up to an identity term in the Hamiltonian. While in general one would need to avoid perturbing the $e$ state (separated from $g$ by $\sim 52$ kHz gap) via transition (ii) in Fig.~\ref{fig_StatePreparation}(a), this requirement may be achieved through symmetry. The laser (ii) matrix element normally vanishes due to the spatial symmetry of the wavefunctions about the lattice site center (see Supplementary Material B). For state preparation, we avoid this by aiming laser (ii) along the $\hat{x}$ direction, which introduces an additional phase to break the symmetry and prevents the integral from vanishing. For rotations, if we aim laser (i) along $\hat{y}$ or $\hat{z}$ instead, transition (ii) will have a vanishing overlap integral, allowing us to generate the desired rotation. For collective rotations such as the $\pi$ pulse in the cluster state protocol, we can even use the same beam as the one used in state preparation.

\subsection{Readout}
The final required piece of the toolbox is to measure the qubits in the decoherence-free subspace. Measurements of the $\hat{\sigma}^{x}$ component can be done by simply reversing the state preparation protocol, then measuring the occupation of the $g$ band. If the qubit was in $\ket{\Rightarrow}$ we would find two atoms, and if in $\ket{\Leftarrow}$, we would find one, offering a simple metric.

The $\hat{\sigma}^{z}$ component is more complex, but can be avoided altogether because we have access to rotations about two axes. We first rotate by $\pi/2$ about the $\hat{x}$ axis on the Bloch sphere, then by $\pi/2$ about the $\hat{z}$ axis, and finally measure $\hat{\sigma}^{x}$. If we were in $\ket{\Uparrow}$ we would end up in $\ket{\Rightarrow}$ and find two atoms, whereas the $\ket{\Downarrow}$ state would be rotated into $\ket{\Leftarrow}$ and yield one.
\section{Conclusions}

We have proposed protocols for using 3D optical lattices populated by atoms in higher bands to realize qubits for quantum computation. Our design embeds the logical states in a decoherence-free subspace, which greatly amplifies the system coherence time while still having scalability, single-site addressability and tunable interactions via Feshbach resonances and electromagnetic field control. State preparation and onsite rotations can be implemented with Raman transitions and external field gradients, allowing for straightforward manipulation of qubit states. Superexchange interactions between the qubits can also be tuned in form and amplitude with field gradients, allowing (among other things) the generation of an Ising model useful for cluster state generation and OTOC measurement.

This system has an abundance of natural extensions that can be implemented without significant modifications to the protocols. While our design allows the creation of a 1D cluster state, 2D states can be generated by reducing the tunneling along $\hat{y}$ and inducing superexchange dynamics along that direction as well (still keeping the lattice depth sufficiently different from $\hat{x}$ so that unwanted band-changing collisions do not occur). Since the dynamics are controlled by a field gradient, we can enable or disable interactions at will, allowing us to build the cluster state up one dimension at a time as the controlled phase gate operations commute. Such a state allows for universal computation using measurement-based protocols. Furthermore, if we seek to study OTOCs, we can use locally-applied fields to induce local perturbations and measure their propagation directly on a site-to-site basis rather than relying upon collective observables.

While our protocols have been described for alkali atoms, straightforward extensions can also be made to systems using alkaline earth atoms instead. The higher-band configuration can be mimicked through the use of different electronic states (with the $g$, $e$ bands replaced by e.g. ground and excited clock states, while the bare spin $\sigma$ remains represented by hyperfine states). The resulting system would still exhibit both inter- and intra-state interactions for the two electronic states. Applying an electronic-state dependent lattice intensity can likewise restrict tunneling to just one of the two electronic states, reproducing our proposed setup. Alkaline earth atoms can provide additional stability and robustness due to their naturally-high coherence times and magnetic field insensitivity. Related prospects, including the use of decoherence-free subspaces, have been explored in Refs.~\cite{scazza2014decoherenceFree,cappellini2014decoherenceFree}.

Other possible applications for our proposed setup include the generation of other useful entangled states such as spin-squeezed states~\cite{ma2011squeezing}, by generating appropriate XXZ-type Hamiltonians~\cite{rey2007squeezing,cappellaro2009squeezing} or similar gap-protected spin interactions~\cite{rey2008squeezing,he2019squeezing}. The restrictions on motion due to the use of higher bands can lead to exotic spatially-correlated physics mimicking spin-orbit coupled systems~\cite{li2016soc}. As a more ambitious goal, we can also explore the fully-resonant regime where $B = U_{\gamma}$ for some $\gamma$ for which nontrivial constrained dynamics can arise on fast timescales~\cite{mamaev2019leapfrog}. Our protocols offer a powerful toolbox for quantum computation and simulation, which is accessible with current state-of-the-art optical lattice experiments, and offers powerful tunability and versatility.

\noindent{\textit{Acknowledgements:}}  This work is supported by the AFOSR grants FA9550-19-1-0275 and FA9550-19-1-7044, the NSF JILA-PFC PHY-1734006 grant, NIST, and NSERC.

\bibliography{TwoBandKondoBibliography}
\bibliographystyle{unsrt}

\clearpage

\onecolumngrid
\renewcommand{\thesubsection}{\Alph{subsection}}
\renewcommand{\appendixtocname}{Supplementary material}
\appendixpageoff
\appendixtitleoff
\begin{appendices}

\begin{center}
\LARGE
Supplementary Material
\normalsize
\end{center}

\section{Fermi-Hubbard derivation and system parameters}
\renewcommand{\thefigure}{A\arabic{figure}}
\renewcommand{\theequation}{A\arabic{equation}}
\setcounter{figure}{0}
\setcounter{equation}{0}
In this appendix, we give a detailed derivation of the Fermi-Hubbard Hamiltonian presented in the main text. While this can be pieced together from textbook sources, we provide the steps here for clarity. The configuration we consider is the ground and first-excited band of a 3D cubic optical lattice, with a single excitation in the $\hat{x}$ direction. Assuming that all lattice depths are high enough to be deep in the Mott insulating regime, the onsite wavefunctions of the states we seek to populate may be written in real space as
\begin{equation}
\begin{aligned}
\label{eq_HigherBandWavefunctions}
\phi_{e}(\vec{r})&= w_1(x) w_0(y) w_0(z)\\
\phi_{g}(\vec{r})&= w_0(x) w_0(y) w_0(z),
\end{aligned}
\end{equation}
where $w_{n}$ is the $n$-th band Wannier function for the corresponding lattice coordinate.

Assuming a deep enough lattice, the dominant s-wave contact interactions in the system will be onsite. In the most general form,
\begin{equation}
\label{eq_GeneralInteraction}
\hat{H}_{\mathrm{U}}=\sum_{j}\sum_{\mu_{1},\mu_{2},\mu_{3},\mu_{4}} \frac{U_{\mu_{1},\mu_{2},\mu_{3},\mu_{4}}}{2}\sum_{\sigma,\sigma'}\hat{c}_{j,\mu_{1},\sigma}^{\dagger}\hat{c}_{j,\mu_2,\sigma'}^{\dagger}\hat{c}_{j,\mu_3,\sigma'}\hat{c}_{j,\mu_4,\sigma},
\end{equation}
where $\hat{c}_{j,\mu,\sigma}$ annihilates a fermion on lattice site $j$, in band state $\mu \in \{e,g\}$, with spin $\sigma \in \{\uparrow,\downarrow\}$, and the interaction coefficient is given by,
\begin{equation}
\label{eq_GeneralInteractionStrength}
U_{\mu_{1},\mu_{2},\mu_{3},\mu_{4}}=g\int d^3 \vec{r} \phi_{\mu_1}^{*}(\vec{r})\phi_{\mu_2}^{*}(\vec{r})\phi_{\mu_3}(\vec{r})\phi_{\mu_4}(\vec{r}),\>\>\>\>\>g=\frac{4\pi \hbar^2 a_s}{m}
\end{equation}
with $g$ the interaction coefficient. Here $m$ is the mass of the atom, and $a_s$ the scattering length.

The dominant terms in Eq.~\eqref{eq_GeneralInteraction} are the intra-band interaction and the inter-band direct and exchange interaction. The $e$-band interactions are set by the $U_{eeee}$ term and can be written as (after cancelling terms and renaming the coefficient),
\begin{equation}
\begin{aligned}
\hat{H}_{\mathrm{U},ee}&=U_{ee}\sum_{j}\hat{n}_{j,e,\uparrow}\hat{n}_{j,e,\downarrow},\\
U_{ee}&=g \int_{-\infty}^{\infty}dx |w_1(x)|^4 \int_{-\infty}^{\infty}dy |w_0(y)|^4 \int_{-\infty}^{\infty}dz |w_0(z)|^4,
\end{aligned}
\end{equation}
with $\hat{n}_{j,\mu,\sigma}=\hat{c}_{j,\mu,\sigma}^{\dagger}\hat{c}_{j,\mu,\sigma}$. Likewise, the $g$ band interactions from $U_{gggg}$ are
\begin{equation}
\begin{aligned}
\hat{H}_{\mathrm{U},gg}&=U_{gg}\sum_{j}\hat{n}_{j,g,\uparrow}\hat{n}_{j,g,\downarrow},\\
U_{gg}&=g \int_{-\infty}^{\infty}dx |w_0(x)|^4 \int_{-\infty}^{\infty}dy |w_0(y)|^4 \int_{-\infty}^{\infty}dz |w_0(z)|^4.
\end{aligned}
\end{equation}
Finally, we have the direct and exchange interactions. They are encapsulated in four terms mixing $e$ and $g$. The combinations are $U_{egeg}$, $U_{egge}$, $U_{geeg}$ and $U_{gege}$. Adding all these terms, evaluating and reducing the expression we arrive at,
\begin{equation}
\begin{aligned}
\hat{H}_{\mathrm{U},eg}&=\frac{U_{eg}}{2}\sum_{j}\left(\hat{n}_{j,e,\uparrow}\hat{n}_{j,g,\downarrow}+\hat{n}_{j,e,\downarrow}\hat{n}_{j,g,\uparrow}\right)-\frac{U_{eg}}{2}\sum_{j}\left(\hat{c}_{j,e,\uparrow}^{\dagger}\hat{c}_{j,e,\downarrow}\hat{c}_{j,g,\downarrow}^{\dagger}\hat{c}_{j,g,\uparrow}+h.c.\right),\\
U_{eg}&=2g\int_{-\infty}^{\infty}dx |w_0(x)|^2 |w_1(x)|^2\int_{-\infty}^{\infty}dy |w_0(y)|^4 \int_{-\infty}^{\infty}dz |w_0(z)|^4.
\end{aligned}
\end{equation}
The first term sets the direct interactions and the second the exchange interactions. Note that we included an uncancelled factor of $2$ in the coefficent, to connect with other conventional derivations.

At this point, we have all the interaction terms, $\hat{H}_{\mathrm{U}}=\hat{H}_{\mathrm{U},ee}+\hat{H}_{\mathrm{U},gg}+\hat{H}_{\mathrm{U},eg}$. The only thing left is the tunneling of atoms. We assume $V_x \ll V_y, V_{z}$ (with $V_{\nu}$ lattice depth along $\hat{\nu}$) so that tunneling is constrained to the $\hat{x}$ direction only. Under this assumption, the only relevant tunneling processes are hopping to nearest-neighbours in the $\hat{x}$ direction. These have corresponding amplitudes of,
\begin{equation}
\begin{aligned}
J_{e,x}&=\int_{-\infty}^{\infty}dx w_{1}^{*}(x)\left[-\frac{\hbar^2}{2m}\frac{d^2}{dx^2}+V_{x}\sin^2\left(\frac{\pi x}{a}\right)\right]w_{1}(x+a),\\
J_{g,x}&=\int_{-\infty}^{\infty}dx w_{0}^{*}(x)\left[-\frac{\hbar^2}{2m}\frac{d^2}{dx^2}+V_{x}\sin^2\left(\frac{\pi x}{a}\right)\right]w_{0}(x+a),
\end{aligned}
\end{equation}
where $a$ is the lattice spacing and $V_{x}\sin^2(\pi x/a)$ the lattice potential (the other components of the Wannier functions integrate out to unity).

Since higher bands have much stronger overlap with neighbouring sites, we make the approximation that $J_{e,x} \gg J_{g,x}$, and assume that only atoms in the $e$-band can tunnel. This yields our tunneling Hamiltonian,
\begin{equation}
\hat{H}_{\mathrm{J}}=-J\sum_{\langle i,j\rangle,\sigma}\left(\hat{c}_{i,e,\sigma}^{\dagger}\hat{c}_{j,e,\sigma}+h.c.\right),
\end{equation}
with $J = -J_{e,x}$. We now have $\hat{H} = \hat{H}_{\mathrm{J}}+\hat{H}_{\mathrm{U}}$, which is Eq.~\eqref{eq_FermiHubbard} in the main text.

One last thing that we do here is give an estimate of the Hamiltonian parameters in the regime of interest. The atom is taken to be $^{40}\mathrm{K}$ with $m = 40$ amu, and the scattering length is set to $a_s = 120 a_0$ via Feshbach resonance ($a_0$ the Bohr radius). The lattice spacing is set to $a = 527$ nm. We assume a lattice of depth $(V_x,V_y,V_z)/E_r=(40,60,60)$, where $E_r = \pi^2 \hbar^2/(2m a^2) = 4.46 \text{ kHz}$ is the recoil energy (in units of $h$). With these parameters, the coefficients evaluate to,
\begin{equation}
J / (2\pi) = 18.9 \text{ Hz}, \>\>\>U_{ee}/(2\pi)=3.38 \text{ kHz}, \>\>\> U_{gg}/(2\pi) = 3.95 \text{ kHz},\>\>\>U_{eg}/(2\pi) = 4.52 \text{ kHz}.
\end{equation}
Note that these interaction strengths are still much less than the band gaps of $\hbar\omega_x\approx 52$ kHz and $\hbar\omega_y=\hbar\omega_z\approx 65$ kHz (with $\omega_{\nu}$ the lattice confinement frequency depending on depth), preventing unwanted scattering. Scattering of the $\hat{x}$ motional excitation into $\hat{y}$, $\hat{z}$ is also inhibited by the gap $\hbar(\omega_y-\omega_x)\approx 13$ kHz due to different lattice depths (see Section E). In units of $J$, these parameters are
\begin{equation}
U_{ee}/J =179,\>\>\>U_{gg}/J = 209,\>\>\>U_{eg}/J = 239.
\end{equation}
From these, the superexchange coefficients may also be found,
\begin{equation}
U_{1}/J = 597,\>\>\>U_{2}/J=119,\>\>\>U_{3}/J=-359.
\end{equation}
These are the parameters we use throughout the plots in the main text.

The dependence of $U_{eg}$ on magnetic and optical fields determines the success of the DFS in rejecting external noise. Since the scattering length is modified near the Feshbach resonance, the effective magnetic moment of the DFS is never zero, but is $>10^2$ lower than that of a single atom when $\gtrsim 5$\,G from the resonance, and $\sim 2 \times 10^{3}$ lower for the $a_s = 120 a_0$ parameter choice above, achieved at $+25$\,G from the resonance. First-order sensitivity to the optical lattice depth can be eliminated by Feshbach tuning $a_S$ to its zero-crossing; the suppression (compared to the single-atom sensitivity) remains $>10^2$ within $\pm$0.6\,G of the zero-crossing, and is $>10$ for the example parameters above. Experimental trials will be necessary to determine which of these external perturbations limit the coherence time of the DFS.

\section{Raman scheme for state preparation}
\renewcommand{\thefigure}{B\arabic{figure}}
\renewcommand{\theequation}{B\arabic{equation}}
\setcounter{figure}{0}
\setcounter{equation}{0}
Here we provide details on the preparation of states in the decoherence-free subspace using a Raman transition, as discussed in the main text [c.f. Fig.~\ref{fig_StatePreparation}(a)]. We start with a band insulator of two atoms per lattice site both in the lowest band $g$, populating the ground hyperfine manifold. For concreteness but without loss of generality we consider the $F=9/2$ states of $^{40}$K atoms. The bare spin states $\sigma \in \{\uparrow,\downarrow\}$ are distinguished by their nuclear-spin projection, associating $\ket{\downarrow}$ with $\ket{m_F=-9/2}$ and $\ket{\uparrow}$ with $\ket{m_F=-7/2}$. The goal is to couple one initially-occupied state in the $g$ band to a desired target state in the $e$ band through an intermediate excited state. A sample intermediate state is a $^{2}P_{1/2}$ electronic state, with nuclear spin $F=7/2$ in the $g$ band, $\ket{^{2}P_{1/2},g;7/2,-7/2}$, where the states are written as $\ket{\Lambda,\mu;F,m_F}$ for electronic state $\Lambda$, band $\mu$, nuclear-spin $F$ and projection $m_F$.

We use a linearly-polarized laser (i) to couple $\ket{^{2}S_{1,2},g;9/2,-7/2}$ with $\ket{^{2}P_{1/2},g;-7/2,-7/2}$, and another linearly-polarized laser (ii) to couple $\ket{^{2}P_{1/2},g;-7/2,-7/2}$ with $\ket{^{2}S_{1,2},e; 9/2,-7/2}$. Large detunings $\Delta$ from the intermediate state ensure it is not populated. A $\pi$ pulse coherently transfers one atom to the state in the $e$ band. The second atom in the initially-occupied $g$ band is unaffected because the excited state has no $m_F = -9/2$ component for laser (i) to couple it to. The result is a state with an $\ket{\uparrow}$ atom in the $e$ band, and a $\ket{\downarrow}$ atom in the $g$ band, corresponding to $\ket{\Rightarrow}$ in the decoherence-free manifold.

The coupling for a single laser between two atomic states $\ket{\mu}$, $\ket{\nu}$ is given by,
\begin{equation}
\label{eq_EffectiveLaserCoupling}
\Omega_{\mu,\nu} = \Omega_{0} C_{q}^{\mu\nu} \int d^{3}\vec{r} \phi_{\mu}^{*}(\vec{r})e^{i \vec{k}_{L}\cdot \vec{r}}\tilde{\phi}_{\nu}(\vec{r}),
\end{equation}
where $\phi_{\mu}$ ($\tilde{\phi}_{\nu}$) is the spatial wavefunction of the band of state $\mu$ ($\nu$) in the $^{2}S_{1/2}$ electronic ground-state ($^{2}P_{1/2}$ excited-state). The two electronic states have different spatial wavefunctions for the same band due to different polarizeabilities. Additionally, $\Omega_{0}$ is the bare drive frequency (set by laser power), $C_{q}^{\mu\nu}=\langle F^{(\mu)}, m_{F}^{(\mu)};1,q|F^{(\nu)}, m_{F}^{(\nu)}\rangle$ is the Clebsch-Gordan coefficient for the transition with $F^{(\mu)}, F^{(\nu)}$ and $m_{F}^{(\mu)}$, $m_F^{(\nu)}$ the total and $\hat{z}$-projected nuclear-spin values for the two coupled states, $q\in \{-1,0,1\}$ is the light polarization (here $q=0$ for both), and $\vec{k}_{L}=\hat{k}_{L}2\pi/\lambda_{L}$ is the laser wavevector (with wavelength $\lambda_{\mathrm{L}}$).

In the appropriate rotating-frame, the Hamiltonian for our desired transition is reduced to a three-level system in the basis of $\{\ket{^{2}S_{1/2},g;9/2,-7/2},\ket{^{2}S_{1/2},e;9/2,-7/2},\ket{^{2}P_{1/2},g;9/2,-7/2}\}$,
\begin{equation}
    \hat{H}_{\mathrm{transfer}}=\left(\begin{array}{ccc}0&0&\Omega_{^{2}S_{1/2}g,^{2}P_{1/2}g}\\
    0&0&\Omega_{^{2}S_{1/2}e,^{2}P_{1/2}g}\\
    \Omega_{^{2}S_{1/2}g,^{2}P_{1/2}g}^{*}&\Omega_{^{2}S_{1/2}e,^{2}P_{1/2}g}^{*}&\Delta\end{array}\right),
\end{equation}
with $\Omega_{^{2}S_{1/2}g,^{2}P_{1/2}g}$ and $\Omega_{^{2}S_{1/2}e,^{2}P_{1/2}g}$ the Rabi frequencies for lasers (i), (ii) respectively. If the detuning $\Delta$ is set to be larger than both off-diagonal terms, as well as any intrinsic decay rate of the $^{2}P_{1/2}$ state to prevent loss due to broadening, we can coherently transfer population from $\ket{^{2}S_{1/2},g;9/2,-7/2}$ to $\ket{^{2}S_{1/2},e;9/2,-7/2}$ state through a second-order process. Should the decay rate be too high, we can instead apply the STIRAP scheme [main text Fig.~\ref{fig_StatePreparation}(b)], for which we would set $\Delta=0$ and instead assign offset Gaussian time-dependence to the off-diagonal couplings.

Note that the excited state's optical-frequency separation is needed for this protocol to work. Normally, laser (ii) would require coupling the $g$ band in $^{2}P_{1/2}$ (with even spatial symmetry along $\hat{x}$) to the $e$ band in $^{2}S_{1/2}$ (with odd spatial symmetry along $\hat{x}$), and their overlap integral would vanish due to an overall odd parity. This is prevented by the position and wavelength-dependent phase factor in the coupling matrix element. The Raman lasers for the chosen transition have wavelengths of $\lambda_{L}\approx 770$ nm, and with a lattice spacing of $a=527$ nm, the wavefunctions vary appreciably over the wavelength. This causes the phase factor $e^{i \vec{k}_{L}\cdot \vec{r}}$ to contribute nontrivially and prevent the integral from vanishing, provided that the laser direction is along $\hat{x}$. Note also that while the spatial wavefunctions will vary significantly between the $^{2}S_{1/2}$ and $^{2}P_{1/2}$ electronic states due to their different polarizeabilities, the extra phase factor should render their integrals nonzero regardless.

\section{Superexchange Hamiltonian derivation}
\renewcommand{\thefigure}{C\arabic{figure}}
\renewcommand{\theequation}{C\arabic{equation}}
\setcounter{figure}{0}
\setcounter{equation}{0}

In this appendix, we consider a general second-order Schrieffer-Wolff transformation to block-diagonalize a Hamiltonian consisting of two state manifolds $\{\ket{m_0}\}$, $\{\ket{m_{\mathrm{V}}}\}$ whose coupling matrix elements are weak compared to the gap between their respective energy scales. The Hamiltonian can be split as,
\begin{equation}
\hat{H}=\hat{H}_0+ \hat{V},
\end{equation}
where $\hat{H}_0$ is assumed to be diagonal, and $\hat{V}$ is a purely off-diagonal perturbation that couples the two manifolds. The diagonal part may be written as,
\begin{equation}
\hat{H}_0 = \hat{P}_0 \hat{H}_{0}\hat{P}_0 + (\mathbbm{1}-\hat{P}_0) \hat{H}_{0}(\mathbbm{1}-\hat{P}_0),
\end{equation}
where $\hat{P}_0 = \sum_{m_0}\ket{m_0}\bra{m_0}$ projects onto the $\{\ket{m_0}\}$ manifold, and $\mathbbm{1}- \hat{P}_0 = \sum_{m_{\mathrm{V}}}\ket{m_{\mathrm{V}}}\bra{m_{\mathrm{V}}}$ onto the $\{\ket{m_{\mathrm{V}}}\}$ manifold. We define the energy gap between the manifolds as,
\begin{equation}
\Delta_g = \min_{m_0,m_{\mathrm{V}}}|E_{m_0}-E_{m_{\mathrm{V}}}|,
\end{equation}
with $E_m$ the diagonal energy $\hat{H}_{0}\ket{m}=E_m\ket{m}$. The large-gap condition required for the perturbation theory's validity corresponds to,
\begin{equation}
\Delta_g \gg ||\hat{V}||,
\end{equation}
with $||.||$ the operator norm. We also assume that the perturbation $\hat{V}$ only has nonzero elements of the form $\ket{m_0}\bra{m_{\mathrm{V}}}$ (i.e. no self-energies or couplings between states in the same manifold).

With the large-gap condition satisfied, cross-manifold transfer of population is inhibited. To lowest-order approximation, the only effect will be a virtual tunneling between states within the same manifold through a second-order process. If only $\{\ket{m_0}\}$ is initially populated, then we can reduce the Hilbert space to just that manifold, and write down an effective perturbation stemming from $\hat{V}$ that couples the $\ket{m_0}$ states. 

To evaluate the effect of the perturbation, we define a superoperator $\mathcal{L}$~\cite{bravyi2011schrieffer},
\begin{equation}
\mathcal{L}\hat{V} = \sum_{m,n}'\frac{\ket{m}\bra{m}\hat{V}\ket{n}\bra{n}}{E_m-E_n},
\end{equation}
where each sum runs over all states $\ket{m} \in \{\ket{m_0}\}\bigcup \{\ket{m_{\mathrm{V}}}\}$, and the prime indicates that we automatically assume the argument is zero if both $\ket{m},\ket{n}$ are from the same manifold. The second-order perturbative contribution is then given by,
\begin{equation}
\label{eq_SuperexchangeDefinition}
\hat{H}_{\mathrm{ex}}=-\frac{1}{2}\hat{P}_{0}\left[\hat{V},\mathcal{L}\hat{V}\right]\hat{P}_{0}.
\end{equation}
One may also rewrite this as,
\begin{equation}
\begin{aligned}
\label{eq_SuperexchangeApp}
    \bra{i}\hat{H}_{\mathrm{ex}}\ket{j} &= \sum_{k}\Delta_{ijk}\bra{i}\hat{H}_{J}\ket{k}\bra{k}\hat{H}_{J}\ket{j},\\
    \Delta_{ijk}&=\frac{1}{2}\left(\frac{1}{E_{i}-E_{k}}+\frac{1}{E_{j}-E_{k}}\right),
\end{aligned}
\end{equation}
for states $\ket{i},\ket{j}\in\{\ket{m_0}\}$ and $\ket{k}\in \{\ket{m_{\mathrm{V}}}\}$, yielding maintext Eq.~\eqref{eq_Superexchange}.

We evaluate this explicitly for our Fermi-Hubbard model [main text Eq.~\eqref{eq_FermiHubbard}], assuming only two neighbouring sites $j$, $j+1$, after which the resulting interaction is extrapolated across the lattice. The unperturbed Hamiltonian is $\hat{H}_{0}=\hat{H}_{\mathrm{U}}+\hat{H}_{\mathrm{B}}$, and the perturbation is $\hat{V}=\hat{H}_{\mathrm{J}}$. The populated manifold $\{\ket{m_{0}}\}$ is defined to be the decoherence-free subspace,
\begin{equation}
\{\ket{m_0}\}=\{\ket{\Uparrow}_{j}\ket{\Uparrow}_{j+1},\ket{\Uparrow}_{j}\ket{\Downarrow}_{j+1},\ket{\Downarrow}_{j}\ket{\Uparrow}_{j+1},\ket{\Downarrow}_{j}\ket{\Downarrow}_{j+1}\},
\end{equation}
with unperturbed energies of,
\begin{equation}
E_{m_0}=\{0,U_{eg},U_{eg},2U_{eg}\}
\end{equation}
The other manifold $\{\ket{m_{\mathrm{V}}}\}$ is written in the Fock basis $\ket{n_{j,e,\uparrow},n_{j,e,\downarrow},n_{j,g,\uparrow},n_{j,g,\downarrow};n_{j+1,e,\uparrow},n_{j+1,e,\downarrow},n_{j+1,g,\uparrow},n_{j+1,g,\downarrow}}$ (semicolon separates the two lattice sites) and consists of,
\begin{equation}
\begin{aligned}
\{\ket{m_{\mathrm{V}}}\}=\{&\ket{1,1,0,1;0,0,1,0},\ket{1,1,1,0;0,0,0,1},\ket{0,0,1,0;1,1,0,1},\ket{0,0,0,1;1,1,1,0}\},
\end{aligned}
\end{equation}
with energies of:
\begin{equation}
E_{m_{\mathrm{V}}}=\{U_{ee}+U_{eg}/2+B/2,\>\>U_{ee}+U_{eg}/2-B/2,\>\>U_{ee}+U_{eg}/2-B/2,\>\>U_{ee}+U_{eg}/2+B/2\}
\end{equation}
The system will tunnel into these states, and back into the decoherence-free manifold through a second-order process. Note that there are also two other possible destination states $\{\ket{1,0,1,0;0,1,0,1},\ket{0,1,0,1;1,0,1,0}\}$ accessible through the second tunneling event, but since these are not initially populated they should have no effect on the dynamics (they would normally be degenerate with one of the decoherence-free states, but the field gradient $B$ shifts their energies away).

Now that we have all of the necessary ingredients, we explicitly compute the effective Hamiltonian corresponding to $\hat{V}$ in the decoherence-free subspace. Evaluating Eq.~\eqref{eq_SuperexchangeApp} for the two sites, we find,
\footnotesize
\begin{equation}
\hat{H}_{\mathrm{ex},2}=\left(
\begin{array}{cccc}
 \frac{2 J^2 U_1}{B^2-U_1^2} & \frac{2 B J^2 \left(B^2-U_1^2+4 U_{ee} U_{eg}\right)}{\left(B^2-U_1^2\right) \left(B^2-U_2^2\right)} &
   -\frac{2 B J^2 \left(B^2-U_1^2+4 U_{ee} U_{eg }\right)}{\left(B^2-U_1^2\right) \left(B^2-U_2^2\right)} & -\frac{2 J^2 U_2
   \left(B^2-U_1 U_3\right)}{\left(B^2-U_1^2\right) \left(B^2-U_3^2\right)} \\
 \frac{2 B J^2 \left(B^2-U_1^2+4 U_{ee} U_{eg }\right)}{\left(B^2-U_1^2\right) \left(B^2-U_2^2\right)} & \frac{2 J^2 U_2}{B^2-U_2^2} &
   -\frac{2 J^2 U_2}{B^2-U_2^2} & -\frac{2 B J^2 \left(B^2-2 U_{eg }^2-U_2 U_3\right)}{\left(B^2-U_2^2\right) \left(B^2-U_3^2\right)} \\
 -\frac{2 B J^2 \left(B^2-U_1^2+4 U_{ee} U_{eg }\right)}{\left(B^2-U_1^2\right) \left(B^2-U_2^2\right)} & -\frac{2 J^2 U_2}{B^2-U_2^2} &
   \frac{2 J^2 U_2}{B^2-U_2^2} & \frac{2 B J^2 \left(B^2-2 U_{eg}^2-U_2 U_3\right)}{\left(B^2-U_2^2\right) \left(B^2-U_3^2\right)} \\
 -\frac{2 J^2 U_2 \left(B^2-U_1 U_3\right)}{\left(B^2-U_1^2\right) \left(B^2-U_3^2\right)} & -\frac{2 B J^2 \left(B^2-2 U_{eg}^2-U_2
   U_3\right)}{\left(B^2-U_2^2\right) \left(B^2-U_3^2\right)} & \frac{2 B J^2 \left(B^2-2 U_{eg}^2-U_2 U_3\right)}{\left(B^2-U_2^2\right)
   \left(B^2-U_3^2\right)} & \frac{2 J^2 U_3}{B^2-U_3^2} \\
\end{array}
\right),
\end{equation}
\normalsize
where $U_{\gamma}$ for $\gamma \in \{1,2,3\}$ are the resonant energy scales in main text Eq.~\eqref{eq_UGamma}. This matrix is used to compute the spin model in main text Eq.~\eqref{eq_SpinModel} by expanding the matrix in a basis of all possible one- and two-point Pauli matrix products. The resulting spin-spin interaction is extrapolated by summing it across the lattice (thus two coupling links for each site in 1D). Note that in all cases, we must also add the matrix elements of $\hat{H}_{0} = \hat{H}_{\mathrm{U}}+\hat{H}_{\mathrm{B}}$ in the decoherence-free subspace directly, as they are unperturbed and contribute dynamics of their own.

\section{$\hat{\sigma}^{x}$ Rotations in the decoherence-free subspace}
\renewcommand{\thefigure}{D\arabic{figure}}
\renewcommand{\theequation}{D\arabic{equation}}
\setcounter{figure}{0}
\setcounter{equation}{0}

We want to implement onsite $\hat{\sigma}^{x}$ rotations for the logic qubits. One way of doing this is by applying an external field gradient, supplementary to the one used for tuning superexchange interactions. The Hamiltonian for a single-site lattice site populated by two atoms (indexed as 1, 2) may be written in real space as~\cite{busch1998twoAtomSingleSite},
\begin{equation}
\hat{H}_{\mathrm{site}} = \hat{H}_{\mathrm{lattice}} + \hat{H}_{\delta B},
\end{equation}
with $\hat{H}_{\mathrm{lattice}}$ given by,
\begin{equation}
\hat{H}_{\mathrm{lattice}}=\left[-\frac{\nabla^2_{1}}{2m}-\frac{\nabla^2_{2}}{2m}+\frac{1}{2}m \omega_{x}^2(x_1^2+x_2^2)+\frac{1}{2}m \omega_{y}^2 (y_1^2+y_2^2)+\frac{1}{2}m \omega_{z}^{2}(z_1^2+z_2^2)+\frac{4\pi \hbar^2 a_s}{m}\delta^{(3)}(\vec{r}_{12})\frac{\partial}{\partial \vec{r}_{12}}\vec{r}_{12}\right]\otimes \mathbbm{1}_{\mathrm{spin}},
\end{equation}
where $(x_1,y_2,z_2)$, $(x_2,y_2,z_2)$ are the positions of atom 1 and 2 respectively, $\vec{r}_{12}=\vec{r}_{1}-\vec{r}_{2}$ is their relative position, $m$ is the atomic mass, the gradient terms are the kinetic energy, the three quadratic terms are the lattice confinement along $\hat{x}$, $\hat{y}$ and $\hat{z}$ (with respective frequencies $\omega_{\nu}$), and $\delta^{(3)}$ is a 3D Dirac delta function. The delta function term is the onsite repulsion with scattering length $a_s$. $\mathbbm{1}_{\mathrm{spin}}$ is the identity operator for the spin degrees of freedom. The field gradient terms are,
\begin{equation}
\hat{H}_{\delta B}= \delta B (\vec{r}_1)\mathbbm{1}_{\mathrm{spatial}}\otimes\hat{\sigma}^{z}_{1}+\delta B (\vec{r}_{2})\mathbbm{1}_{\mathrm{spatial}}\otimes\hat{\sigma}^{z}_{2},
\end{equation}
with $\hat{\sigma}^{z}_{1}$, $\hat{\sigma}_{2}^{z}$ the Pauli operators for the first and second spin, $\mathbbm{1}_{\mathrm{spatial}}$ the identity operator for the spatial degrees of freedom, and $\delta B(\vec{r})$ describing the spatial dependence of the field gradient that we choose.

Our decoherence-free subspace state wavefunctions are written as [c.f. Eq.~\eqref{eq_HigherBandWavefunctions}],
\begin{equation}
\begin{aligned}
\ket{\Uparrow}&\equiv \frac{1}{2}\left(\ket{\uparrow,\downarrow}+\ket{\downarrow,\uparrow}\right)\left[\phi_{e}(\vec{r}_{1})\phi_{g}(\vec{r}_{2})-\phi_{g}(\vec{r}_{1})\phi_{e}(\vec{r}_{2})\right],\\
\ket{\Downarrow}&\equiv \frac{1}{2}\left(\ket{\uparrow,\downarrow}-\ket{\downarrow,\uparrow}\right)\left[\phi_{e}(\vec{r}_{1})\phi_{g}(\vec{r}_{2})+\phi_{g}(\vec{r}_{1})\phi_{e}(\vec{r}_{2})\right].
\end{aligned}
\end{equation}
For the sake of obtaining an analytic result, the Wannier functions can be approximated by harmonic oscillator wavefunctions, which is a fair description for deep lattices,
\begin{equation}
\begin{aligned}
w_0(\nu) &\approx \left(\frac{m \omega_{\nu}}{\pi \hbar}\right)^{1/4}e^{-\frac{m \omega_{\nu}\nu^2}{2 \hbar}},\\
w_1(\nu) &\approx \frac{\sqrt{2}}{\pi^{1/4}}\left(\frac{m \omega_{\nu}}{\hbar}\right)^{3/4}\nu\> e^{-\frac{m \omega_{\nu}\nu^2}{2 \hbar}}.
\end{aligned}
\end{equation}
Assuming that atoms are restricted to the two-state subspace $\{\ket{\Uparrow},\ket{\Downarrow}\}$, we project the Hamiltonian into it, yielding a $2 \times 2$ matrix. All terms except $\hat{H}_{\delta B}$ do not contain an explicit spin component, and will thus be diagonal (conversely, $\hat{H}_{\delta B}$ turns out to be purely off-diagonal). We calculate the diagonal terms explicitly by integrating over the spatial components,
\begin{equation}
\begin{aligned}
\langle \Uparrow|\hat{H}_{\mathrm{site}}|\Uparrow\rangle &= \frac{1}{2} \int d^{3}\vec{r}_{1} \int d^{3}\vec{r}_{2} \left[\phi_{e}(\vec{r}_{1})\phi_{g}(\vec{r}_{2})-\phi_{g}(\vec{r}_{1})\phi_{e}(\vec{r}_{2})\right] \cdot \hat{H}_{\mathrm{lattice}}\cdot\left[\phi_{e}(\vec{r}_{1})\phi_{g}(\vec{r}_{2})-\phi_{g}(\vec{r}_{1})\phi_{e}(\vec{r}_{2})\right]\\
&=\hbar (2 \omega_x + \omega_y + \omega_z).
\end{aligned}
\end{equation}
These are just the total band energies for the two atoms. The other diagonal element is computed likewise; unlike the previous, the spatial symmetry causes it to be affected by interactions,
\begin{equation}
\begin{aligned}
\langle \Downarrow|\hat{H}_{\mathrm{site}}|\Downarrow\rangle &= \frac{1}{2} \int d^{3}\vec{r}_{1} \int d^{3}\vec{r}_{2} \left[\phi_{e}(\vec{r}_{1})\phi_{g}(\vec{r}_{2})-\phi_{g}(\vec{r}_{1})\phi_{e}(\vec{r}_{2})\right]\cdot \hat{H}_{\mathrm{lattice}} \cdot \left[\phi_{e}(\vec{r}_{1})\phi_{g}(\vec{r}_{2})-\phi_{g}(\vec{r}_{1})\phi_{e}(\vec{r}_{2})\right]\\
&=\hbar (2 \omega_x + \omega_y + \omega_z) + a_s \sqrt{\frac{2 m \hbar \omega_x \omega_y \omega_z}{\pi }}=\hbar (2 \omega_x + \omega_y + \omega_z)+U_{eg}.
\end{aligned}
\end{equation}
The $a_s$-proportional term is equal to $U_{eg}$, the exchange interaction energy.

We now turn to the off-diagonal elements rising from the $\delta B$ terms. The spin part of the Hamiltonian for the two particles can be written in the bare spin basis $\{\ket{\uparrow,\uparrow}, \ket{\uparrow,\downarrow},\ket{\downarrow,\uparrow},\ket{\downarrow,\downarrow}\}$ as,
\begin{equation}
    \hat{H}_{\delta B}=\left(\begin{array}{cccc}\delta B(\vec{r}_{1})+\delta B (\vec{r}_{2})&0&0&0\\
    0&\delta B (\vec{r}_{1})-\delta B (\vec{r}_{2})&0&0\\0&0&-\delta B (\vec{r}_{1})+\delta B (\vec{r}_{2})\\0&0&0&-\delta B (\vec{r}_{1})-\delta B (\vec{r}_{2})\end{array}\right).
\end{equation}
We project it into the decoherence-free subspace via,
\begin{equation}
    \hat{P}\hat{H}_{\delta B}\hat{P}=\left(\begin{array}{cccc}0&0&0&0\\0&0&\delta B(\vec{r}_{1})-\delta B (\vec{r}_{2})&0\\0&\delta B(\vec{r}_{1})-\delta B (\vec{r}_{2})&0&0\\0&0&0&0\end{array}\right),\>\>\>\hat{P}=\frac{1}{\sqrt{2}}\left(\begin{array}{cccc}0&0&0&0\\0&1&1&0\\0&1&-1&0\\0&0&0&0\end{array}\right),
\end{equation}
giving rise to an off-diagonal coupling between the $\ket{\Uparrow},\ket{\Downarrow}$ states equal to $\delta B(\vec{r}_{1})-\delta B (\vec{r}_{2})$. Integrating over the spatial components, we deduce that the field gradient terms in the decoherence-free subspace are
\begin{equation}
\begin{aligned}
\langle \Uparrow |\hat{H}_{\mathrm{site}}| \Downarrow\rangle = \frac{1}{2} \int d^{3}\vec{r}_{1} \int d^{3}\vec{r}_{2} \left[\phi_{e}(\vec{r}_{1})\phi_{g}(\vec{r}_{2})-\phi_{g}(\vec{r}_{1})\phi_{e}(\vec{r}_{2})\right]&\cdot \left[\delta B(\vec{r}_1) - \delta B (\vec{r}_2)\right]\\
&\cdot\left[\phi_{e}(\vec{r}_{1})\phi_{g}(\vec{r}_{2})+\phi_{g}(\vec{r}_{1})\phi_{e}(\vec{r}_{2})\right].
\end{aligned}
\end{equation}
The exact value depends on the spatial dependence we choose for the gradient. If we assume a \textit{linear} gradient, the above integral evaluates to zero. While we used a linear gradient for cross-site interactions, here the matrix element cancels out because of the wavefunction symmetries across the lattice site. This is a good thing, as otherwise our linear gradient $B$ for tuning the interactions would create unwanted onsite effects. We instead use a \textit{quadratic} gradient,
\begin{equation}
\delta B (\vec{r}) = \delta\Omega (x-x_0)^2,
\end{equation}
where $\delta\Omega$ is the gradient strength (in units of energy per length squared), and the direction of the gradient is chosen to be along $\hat{x}$ since our motionally-excited band is along that direction (otherwise the integral will vanish). We also assume a possible offset $x_0$, although this will not change the result. Evaluating the integral yields,
\begin{equation}
\langle \Uparrow |\hat{H}_{\mathrm{site}}| \Downarrow\rangle = \frac{\hbar \> \delta\Omega}{m \omega_x}.
\end{equation}
Our Hamiltonian in the decoherence-free subspace is then given by
\begin{equation}
\hat{H}_{\mathrm{site}} = \left(\begin{array}{cc} \hbar (2 \omega_x + \omega_y + \omega_z) &\frac{\hbar \>\delta\Omega}{m \omega_x} \\\frac{\hbar \>\delta\Omega}{m \omega_x}& \hbar (2 \omega_x + \omega_y + \omega_z) + U_{eg}\end{array}\right).
\end{equation}
The diagonal terms $\hbar (2 \omega_x + \omega_y + \omega_z)$ from the band energies impose a constant energy shift and can be ignored. Dropping them gives Eq.~\eqref{eq_HamiltonianDecoherenceFreeWithGradient} in the main text. Note that the lack of dependence on $x_0$ occurs because the relevant cross term is linear in either $x_1$ or $x_2$, causing its contribution to vanish due to symmetry.

To do $\hat{\sigma}^{x}$-type rotations with this gradient requires the off-diagonal term to be much larger than the diagonal interaction term,
\begin{equation}
\frac{\hbar\>\delta\Omega}{m \omega_x} \gg U_{eg}.
\end{equation}
This can be accomplished by turning off $U_{eg}$ via Feshbach resonance, or with a strong enough gradient. For the parameters at the end of Section A, the exchange interaction is approximately $U_{eg} \approx 4.5$ kHz, meaning that we would need $\delta\Omega \gtrsim 10^{6} \text{ Hz}/\mu \text{m}^2$ for a comparable off-diagonal coupling. Such a requirement is stringent, and it would be easier to just turn off the interactions. However, it also shows robustness to unwanted field curvature. The decoherence-free subspace is implicitly immune to constant fields, and we already showed that linear gradients will have no on-site effect. If any present quadratic gradients are much smaller than the above requirement, we find that the three lowest-order Taylor expansion terms of any magnetic field fluctuation do not affect the decoherence-free subspace.

\section{Robustness to scattering into other bands}
\renewcommand{\thefigure}{E\arabic{figure}}
\renewcommand{\theequation}{E\arabic{equation}}
\setcounter{figure}{0}
\setcounter{equation}{0}

A common limitation of working with higher bands is the system's vulnerability to unwanted band-changing collisions. Sufficiently large interactions can enable two-atom processes that will move both atoms out of the decoherence-free subspace and into other excited bands.

To verify the robustness of our subspace, we simulate the dynamics of a single lattice site, including the other states with one band excitation along $\hat{y}$ and $\hat{z}$ with respective spatial wavefunctions $w_0(x)w_1(y)w_0(z)$ and $w_0(x)w_0(y)w_1(z)$. We label all the excited bands $e_{\nu}$ for $\nu \in \{x,y,z\}$. Higher second-excited bands are not included because their population would require creating an additional motional excitation with energy $\hbar\omega_{\nu}$, which would cost $\approx 52$ kHz even for the shallower $\hat{x}$ direction (compared to interaction energy scales of $\sim 5$ kHz).  The interactions are given by Eqs.~\eqref{eq_GeneralInteraction},~\eqref{eq_GeneralInteractionStrength} with only a single site $j=1$,
\begin{equation}
\hat{H}_{\mathrm{U}}=\sum_{\mu_{1},\mu_{2},\mu_{3},\mu_{4}} \frac{U_{\mu_{1},\mu_{2},\mu_{3},\mu_{4}}}{2}\sum_{\sigma,\sigma'}\hat{c}_{\mu_{1},\sigma}^{\dagger}\hat{c}_{\mu_2,\sigma'}^{\dagger}\hat{c}_{\mu_3,\sigma'}\hat{c}_{\mu_4,\sigma},
\end{equation}
We also include an onsite magnetic field $\hat{H}_{B}$,
\begin{equation}
    \hat{H}_{B}=\frac{B}{2}\sum_{\mu}\left(\hat{n}_{\mu,\uparrow}-\hat{n}_{\mu,\downarrow}\right),
\end{equation}
and band energy shifts,
\begin{equation}
    \hat{H}_{\mathrm{band}}=\frac{1}{2}\sum_{\mu}\sum_{\nu}\left(1+2m_{\mu,\nu}\right)\hbar\omega_{\nu}\left(\hat{n}_{\mu,\uparrow}+\hat{n}_{\mu,\downarrow}\right),
\end{equation}
where $m_{\mu,\nu}$ is the excitation number of band $\mu \in \{g,e_x,e_y,e_z\}$ along direction $\nu \in \{x,y,z\}$ (i.e. $m_{\mu,\nu}=1$ if $\mu$ is excited along $\nu$, and 0 otherwise). We use the standard parameters from the end of Section A. The system is prepared in $\ket{\Rightarrow}=\ket{1,0,0,1,0,0,0,0}$, where the Fock ordering is $\ket{n_{g,\uparrow},n_{g,\downarrow},n_{e_x,\uparrow},n_{e_x,\downarrow},n_{e_y,\uparrow},n_{e_y,\downarrow},n_{e_z,\uparrow},n_{e_z,\downarrow}}$.

\begin{figure}
\centering
\includegraphics[width=0.8\linewidth]{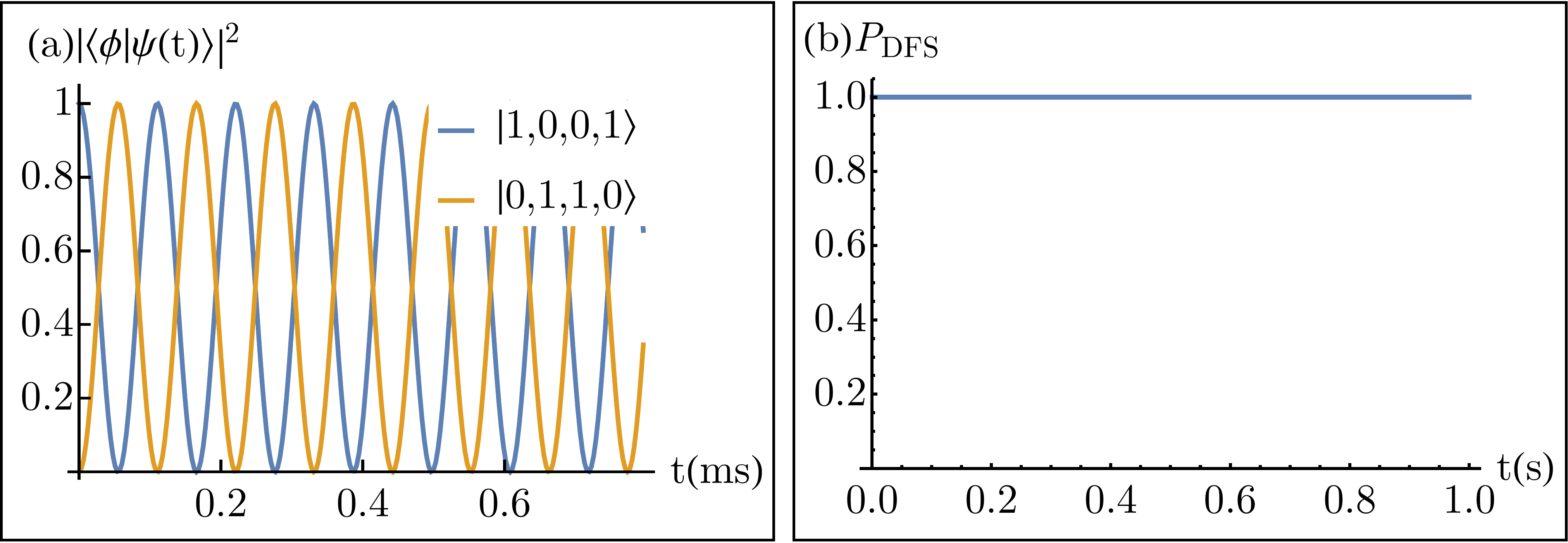}
\caption{(a) Coherent oscillations of population in the decoherence-free states $\ket{\phi} \in \{\ket{\Rightarrow},\ket{\Leftarrow}\}$ for a single site, with $\ket{\psi(t)}$ the full wavefunction including other singly-excited bands in the Hilbert space. (b) Total population in the decoherence-free subspace (sum of the two components in the previous panel) to longer times. For comparison, the cluster state generation time is $t_c \sim 100-300$ ms.}
\label{fig_HigherBand}
\end{figure}

The dynamics of this system are shown in Fig.~\ref{fig_HigherBand}. Panel (a) depicts the coherent oscillations between $\ket{\Rightarrow}$ and $\ket{\Leftarrow}=\ket{0,1,1,0,0,0,0,0}$ rising from the exchange interaction $U_{eg}$. Panel (b) shows that no population leaks outside the subspace defined by these two states, plotting the total overlap $P_{\mathrm{DFS}}=|\bra{\Rightarrow}\psi(t)\rangle|^2+|\bra{\Leftarrow}\psi(t)\rangle|^2$ for the timescales.

Note that while having multiple lattice sites may broaden the bands, the tunneling is sufficiently weak compared to the interactions and bandgaps that it should not induce band transitions. We are already using second-order perturbative effects between the lattice sites to generate our spin model; any unwanted effects would have an additional energy of $\sim 13$ kHz in the perturbation theory denominators (compared to the numerator depending on tunneling rate $\sim 20$ Hz) and would thus be negligible.

\end{appendices}

\end{document}